\documentclass{emulateapj}

\usepackage{graphicx}
\usepackage{epsfig}
\usepackage{epstopdf}
\usepackage{amsmath}
\usepackage{rotating}

\def\gs{\mathrel{\raise0.35ex\hbox{$\scriptstyle >$}\kern-0.6em
\lower0.40ex\hbox{{$\scriptstyle \sim$}}}}
\def\ls{\mathrel{\raise0.35ex\hbox{$\scriptstyle <$}\kern-0.6em
\lower0.40ex\hbox{{$\scriptstyle \sim$}}}}

\newcommand{\um}{\,$\mu$m}
\newcommand{\umm}{\,$\mu$m }

\newcommand{\spitz}{{\sl Spitzer}}
\newcommand{\spitzz}{{\sl Spitzer }}

\newcommand{\msun}{\,$\rm{M}_{\odot}$}
\newcommand{\asec}{\,$^{\prime\prime}$\,}
\newcommand{\amin}{\,$^{\prime}$\,}

\begin{document}

\title{Star formation and dust obscuration in the tidally distorted galaxy NGC 2442}

\author{Anna Pancoast$^{1,2}$, Anna Sajina$^1$, Mark Lacy$^3$, Alberto Noriega-Crespo$^4$, and Jeonghee Rho$^5$}

\affil{$^1$ Haverford College, Haverford, PA, 19041}
\affil{$^2$ UC Santa Barbara, Santa Barbara, CA, 93106}
\affil{$^3$ North American ALMA Science Center, National Radio Astronomy Observatory, Charlottesville, VA, 22903}
\affil{$^4$ Spitzer Science Center, Caltech, Pasadena, CA, 91125}
\affil{$^5$ NASA Ames Research Center, SOFIA, M/C 211-3, Moffett Field, CA, 94035}


\begin{abstract}
We present a detailed investigation of the morphological distribution and level of star formation and dust obscuration in the nearby tidally distorted galaxy NGC2442.  \spitzz images in the IR at 3.6, 4.5, 5.8, 8.0\,\um, and 24\um\  and {\sl GALEX} images at 1500\AA{} and 2300\AA{} allow us to resolve the galaxy on scales between $\sim$240-600\,pc.  We supplement these with archival data in the B, J, H, and K bands. We use the 8\um, 24\um\ and FUV (1500\AA) emission to study the star formation rate (SFR). We find that globally, these tracers of star formation give a range of results of $\sim$\,6\,--\,11\msun/yr, with the dust-corrected FUV giving the highest value of SFR.  We can reconcile the UV and IR-based estimates by adopting a steeper UV extinction curve that lies in between the starburst (Calzetti) and SMC extinction curves. However, the regions of highest SFR intensity along the spiral arms are consistent with a starburst-like extinction. Overall, the level of star-formation we find is higher than previously published for this galaxy, by about a factor of two, which, contrary to previous conclusions, implies that the interaction that caused the distorted morphology of NGC2442 likely also triggered increased levels of star-formation activity. We also find marked asymmetry in that the north spiral arm has a noticeably higher SFR than the southern arm. The tip of the southern spiral arm shows a likely tidally-distorted peculiar morphology. It is UV-bright and shows unusual IRAC colors, consistent with other published tidal features IRAC data.  Outside of the spiral arms, we discover what appears to be a superbubble, $\sim$\,1.7\,kpc across, which is seen most clearly in the IRAC images. Significant H${\alpha}$, UV and IR emission in the area also suggest vigorous ongoing star-formation.  A known, recent supernova (SN1999ga) is located at the edge of this superbubble. Although speculative at this stage, this area suggests a large star-forming region with a morphology shaped by generations of supernovae.  Lastly, we discover an 8\um\ (PAH) circumnuclear ring with an $\sim$\,0.8\,kpc radius. The H$\alpha$ emission is largely concentrated inside that ring and shows a vague spiral structure in the rest of the galaxy.  The nuclear region shows the highest obscuration levels in the galaxy (A$_{1600}$\,$\sim$\,3\,--\,4) most likely due to the circumnuclear dust ring. 
\end{abstract}


\section{Introduction}
One of the most important quantities we derive for both individual galaxies and for populations used in studies of galaxy evolution is the star formation rate (SFR). Historically, we have relied on two principle diagnostics: the UV emission of young stars, and the IR emission of starlight-heated dust. The principal drawback of the UV approach is the strong extinction dependence, which can be somewhat corrected, for example, by using the UV slope \citep{calzetti94,meurer,kong04}. This relation is different for non-starbursting vs. starbursting galaxies potentially due to the influence of older stellar populations \citep{cortese06,dale07}, as well as differences in extinction laws \citep{boquien209}.  The relation often fails for IR-selected starbursts \citep{buat05}, including sub-mm galaxies (SMGs) and dusty red galaxies (DRGs) \citep[see e.g.][]{reddy06}.  The IR approach's principle drawback is SED variations which can affect bolometric vs. monochromatic estimates, as well as dust heating by older stellar populations. This is believed to be especially significant for PAH emission (e.g. broadband 8\,\um) and long wavelength cold dust emission, but less of an issue for the warm dust dominating the mid-IR continuum \citep[e.g. broadband 24\,\um;][]{peeters02, calzetti}.  IR-based relations may also underestimate the SFR if a significant fraction of the starlight is unobscured by dust, although this is partially mitigated by the conversion factors. More fundamentally, the conversion relations we use are ultimately calibrated on stellar population synthesis models and nearby populations of both normal and starburst galaxies \citep[see][]{kennicutt98} typically using total galaxy values.  This effectively averages highly obscured with unobscured regions and could lead to underestimating the dust content in high redshift interacting galaxies \citep{charmandaris04}. In order to further understand the limitations of these estimates, we would like to see how well these relations hold-up on smaller galactic scales, as was done in the case of M51 by \citet{calzetti}. 

By $z$\,$\sim$\,1, the star-formation rate density is believed to be dominated by luminous infrared galaxies or LIRGs, \citep{lefloch05}. Recent work suggests that in such galaxies star-formation activity is likely to be largely triggered by mergers and tidal interactions \citep{bridge07,shi09}. Therefore, we are particularly interested to test how different SFR tracers and broadband obscuration diagnostics behave in galaxies undergoing such interactions. Moreover, an investigation of the localized SFR in such galaxies can give us insight into the details of how interactions trigger star-formation. 

The nearby peculiar spiral galaxy NGC 2442 provides such an opportunity. It appears tidally distorted with areas of high star formation rate in the asymmetric spiral arms and nuclear region.  The tidal distortion is thought to be the result of either ram-pressure stripping from interaction with an inhomogeneous IGM \citep{ryder01} or a tidal interaction with a neighboring galaxy \citep{mihos}. At a distance of 20.7\,Mpc ($z$\,=\,0.004890, NED\footnote{NASA/IPAC Extragalactic Database}), NGC 2442 is significantly closer to us than, for example, the bulk of the members of the Arp Atlas of peculiar galaxies \citep{arp}\footnote{While the original atlas does not include distance estimates, \citet{smith} have selected a sub-set of the Arp galaxies which have angular sizes $>$3$^{\prime}$ and hence are biased to the more nearby members of the Arp catalog. This sub-sample has luminosity distances in the range $\sim$\,6\,--\,150\,Mpc, with only two galaxies having smaller distances than NGC2442.}.  Some interesting observations include: a LINER nucleus \citep{bajaja},  an anomalous magnetic region somewhat offset from the bulk of the galaxy \citep{harnett}, a recent supernova,  SN 1999ga \citep{woodings99,pastorello}, and lastly a massive dark H{\sc i} cloud just beyond the northern spiral arm \citep{ryder01}.  

In this paper we use a multi-wavelength dataset to identify and compare areas of high star-formation rate and dust obscuration in NGC 2442.  We identify regions of high star-formation rate using emission in the UV and infrared: spectral regions that trace the original and reprocessed UV light from young, hot stars, respectively \citep[for a review of star formation tracers see][]{kennicutt98}.  The Galaxy Evolution Explorer \citep[{\sl GALEX};][]{martin} and \spitzz Space Telescope \citep{werner} allow for high sensitivity observations of NGC 2442 in the UV and IR.  Throughout this paper we assume a flat $\Lambda$-dominated Universe with $\rm{H_{o}=71\,km\,s^{-1}\,Mpc^{-1}}$, $\Omega_{\rm{M}}=0.27$, and $\Omega_{\Lambda}=0.73$ \citep{spergel}.  At the redshift of NGC2442, this implies 1\asec\,$\sim$\,100\,pc. 

\section{Data}

\subsection{{\sl Spitzer} mid-IR images}
NGC 2442 was observed at 3.6, 4.5, 5.8, and 8.0\um\ with the \spitz\ Infrared Array Camera \citep[IRAC;][]{fazio} and at 24\um\ with the Multiband Imaging Photometer for \spitz\ \citep[MIPS;][]{rieke} as part of the {\sl Spitzer} science verification observations.  The IRAC images were taken on November 21, 2003.   The MIPS data were also obtained early in the mission to cross-check the performace of the 70\um\, array in all observing modes, including Scan Mapping, that takes data on the 3 MIPS bands (24, 70 \& 160\um). The scan data was taken at a slow scan rate (10 seconds per pointing) over a half degree leg (FOV $\sim 5\arcmin~\times 30\arcmin$),  with a total integration time of 50 secs at 24\um. In this paper, we mostly focus on the 24\um\ data due to its higher resolution; however, we refer to the 70\um\ and 160\um\ data as well when discussing the total SFR. The IRAC images ideally resolve the galaxy on scales of 1.0\asec, 1.1\asec, 1.4\asec, and 2.0\asec\ for the 3.6\um\ through 8.0\um\ images respectively. However, the effective resolution of IRAC channels 1 and 2 is limited to $~$1.2\asec\ due to the detector pixelization.

The 24\um\ image has a resolution of 5.8\asec.  All {\sl Spitzer} images are in units of MJy/sr, after all the standard pipeline processing. For our analysis, we convert all other images (see below) to the same units.  We also perform a background subtraction for each image by subtracting the mean of the background in a large patch near the galaxy (the mean and median were essentially the same).  The 5.8\,\umm image has a patchy background (due to varying dark levels in the detector as a function of time -- the ``first frame" effect), but for a bright galaxy like ours and given that the 5.8\um\ band is not crucial in our analysis, we did not consider it necessary to clean this image further. For this image, we used the average value from multiple patches to perform a background subtraction.  

\subsection{Optical/near-IR images}
An optical B-band (0.44\,\um) image of NGC 2442 was taken using the Cerro Tololo Inter-American Observatory (CTIO) as part of the Bright Spiral Galaxy Survey (Eskridge et al. 2002) and was available through the NASA/IPAC Extragalactic Database (NED).  The B-band image has a resolution of 1.5\asec\ and was first background subtracted and then flux-calibrated using the value for the total galaxy flux given in \citet{sersic}. This is a very rough `calibration', but as we only make very limited use of this image, we felt it was sufficient. 

Near-IR images of NGC 2442 in the J, H, and K bands (1.2, 1.65, 2.2\,\um) were obtained as part of the Two Micron All Sky Survey \citep[2MASS;][]{skrutskie} and available, already background subtracted, through NED \citep[for details see][]{jarrett}.  We converted the units of these images from pixel counts to MJy/sr by relating flux to pixel counts by:
\begin{equation}
\rm{\frac{F[Jy]}{pixel} = F_o*10^{-0.4(MAGZP-2.5\log(S))}}
\end{equation}
where {\sc magzp} is given in the FITS header for each image, and $\rm{F_o[J]=1594\,Jy}$, $\rm{F_o[H]=1024\,Jy}$, and $\rm{F_o[K]=666.8\,Jy}$, as given by \citet{cohen}.  The 2MASS images have a resolution of 2.5\asec.

\subsection{UV images}
{\sl GALEX} images of NGC 2442 were taken in the far-ultraviolet around 1500\AA{} (FUV, 1350-1750\AA{}) and in the near-ultraviolet around 2300\AA{} (NUV, 1750-2750\AA{}) as part of targeted observations (PI: Mark Lacy).  The NUV image was taken on March 24, 2005 with an exposure time of 2.64\,hours and the FUV image was taken on January 24, 2006 with an exposure time of 1.54\,hours.  These images were included in the UV Atlas of Nearby Galaxies by \citet{gil}.  The {\sl GALEX} images have resolutions of 4.2\asec\ for the FUV and 5.3\asec\ for the NUV \citep{morrissey}.  We converted the units of these images from cps (counts per second) to MJy/sr using $\rm{F[MJy/sr]}=\rm{0.638*F_{NUV}[cps]/pix}$ and $\rm{F[MJy/sr]}=\rm{2.036*F_{FUV}[cps]/pix}$ which are derived from:
\begin{equation}
\rm{m_{AB} = -2.5*\log(\frac{cps}{rr})+m_{o,AB}}
\end{equation}
where $\rm{rr\sim1}$, $\rm{m_{o,AB}[NUV]=20.08}$, and $\rm{m_{o,AB}[FUV]=18.82}$ (see the {\sl GALEX} Observer's Guide).  We background subtract both images and check that the final total fluxes for the galaxy are very close to those calculated by \citet{gil}.  

\subsection{Foreground extinction correction}
This galaxy has significant foreground extinction, which strongly affects the observed UV emission, but also affects the B, J, H, and K bands while, by extrapolation, is likely negligible in the IRAC bands.
We adopt the following values for the UV foreground: $\rm{A_{FUV}=1.60}$ and $\rm{A_{NUV}=1.62}$ \citep{gil}. The optical/near-IR MW extinction values, taken from NED, are $\rm{A_{B}=0.874}$, $\rm{A_{J}=0.183}$, $\rm{A_{H}=0.117}$, and $\rm{A_{K}=0.074}$. These corrections are applied to all UV-to-near-IR images prior to any subsequent analysis.

\section{Analysis}

\subsection{Image Alignment}
To facilitate the multi-wavelength analysis presented in this paper, the above images were aligned and rebinned to a common pixel scale. This procedure was carried out using the IDL {\sc frebin} function, which conserves flux. The pixel scale was chosen to be at least as large as the resolution of the poorest resolution image.  For analysis that only uses IRAC and 2MASS images, we adopt a pixel scale of 2.4\asec, and for the rest we adopt a pixel scale of 6.0\asec.  The first pixel scale corresponds to  240\,pc, and the second to 600\,pc at the redshift of NGC 2442.  

\begin{figure}[h!]
\begin{center}
\includegraphics[scale=0.5]{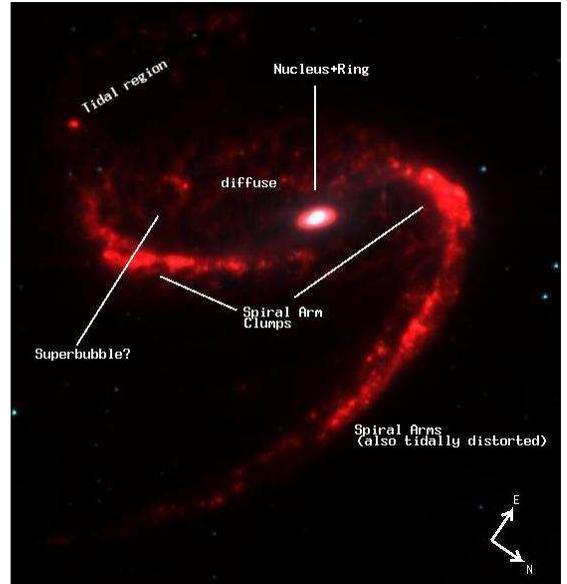}
\includegraphics[scale=0.85]{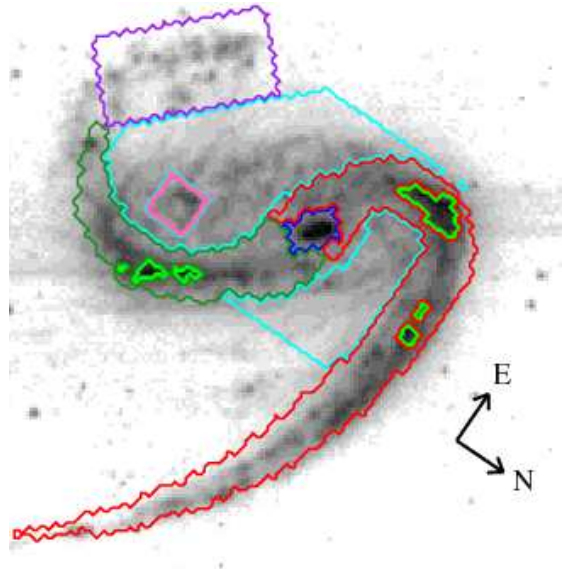}
\caption{{\it Top: } The IRAC color image of NGC2442 ({\it blue} 3.6\um; {\it green} 4.5\um; and {\it red} 8\um) where we label the key regions of interest. {\it Bottom:} A 7\amin\,$\times$\,7\amin\ cutout of the 8$\mu$m image of NGC 2442. The colored contours show the masks applied in this paper to select regions of interest, where red shows the north spiral arm mask, dark green shows the south spiral arm, green shows the spiral clumps mask, cyan shows the diffuse region mask, pink shows the superbubble mask, dark blue shows the nuclear region, and magenta shows the southern tidal region mask. Note that the nuclear region mask includes both the nucleus and nuclear ring (see left). \label{fig_masks}}  
\end{center}
\end{figure}

\subsection{Isolating Regions of Interest \label{sect_masks}}
In order to compare star formation and dust obscuration in different areas of NGC 2442, we generate masks that isolate prominent physical features of the galaxy as shown in Figure \ref{fig_masks}:
\begin{itemize}
\item{\it North and south spiral arms:} These are defined as shown in Figure\,\ref{fig_masks}. Previous work by \citet{mihos}, based on H$\alpha$ observations showed that the northern spiral arm is likely stronger in terms of star-formation compared with the southern arm and hence we also include this north-south distinction. Note that the tip of the southern arm is treated separately in the ``Tidal region'' below.
\item{\it Star-forming clumps/knots:} These are areas inside the spiral arms that are particularly bright at 24\um\ (presumably due to intense star-formation). We define the maks, by setting a threshold of 24\um\ pixel-value of $>$\,7.5\,MJy/ster which corresponds to a star formation rate intensity of 0.062\msun/yr$\rm{/kpc^2}$.
\item{\it Nuclear region:}
This region can be resolved into a nucleus and a circumnuclear ring in the IRAC images (see Figure \ref{fig_masks}), which was also suggested by \citet{mihos} on the basis of their H$\alpha$ map of the galaxy. The radius of this circumnuclear ring is $\sim$\,0.8\,kpc. It is most prominent in the IRAC 8\um\ image, suggesting it is a strong PAH emitter. As mentioned in the introduction, the nucleus of NGC2442 has a LINER spectrum \citep{bajaja}. Note that, apart from its treatment in the body of this paper alongside the rest of the regions of interest, we include an appendix where we discuss the observed properties of the nuclear region in further detail.  
\item{\it Diffuse:} This includes the filamentary structure outside the spiral arms in the IRAC images, especially prominent on the southern side of the galaxy (see Figure~1). This region is also a strong UV emitter.
\item{\it Tidal region:} This may simply be an extension of the southern arm, but due to its disturbed morphology, suggestive of tidal debris,  we treat it separately. 
\item{\it Superbubble:} This is a spherical-looking area within the diffuse region, to the south-west of the nuclear region, with a diameter of $\rm{\sim1.7\,kpc}$. SN 1999ga \citep{woodings99,pastorello} is found on the edge of this structure. 
\end{itemize}
Similar studies of resolved galaxies have also used multi-pixel apertures placed specifically over regions of interest \citep[see][]{calzetti, smith, boquien09, zhang09}. Our approach means that essentially all of the galaxy falls inside one of our masks, and therefore we can truly investigate, for example, which region dominates which part of the total spectral energy distribution (SED). The total galaxy aperture covers $\rm{1063\,kpc^2}$ and includes all the regions of interest. Table\,1 lists the different masks (including the total galaxy mask) in order of decreasing area covered. We also list the total luminosities in each region in the FUV, 8\um, and 24\um\ bands.  

\begin{figure}[h!]
\begin{center}
\includegraphics[scale=0.6]{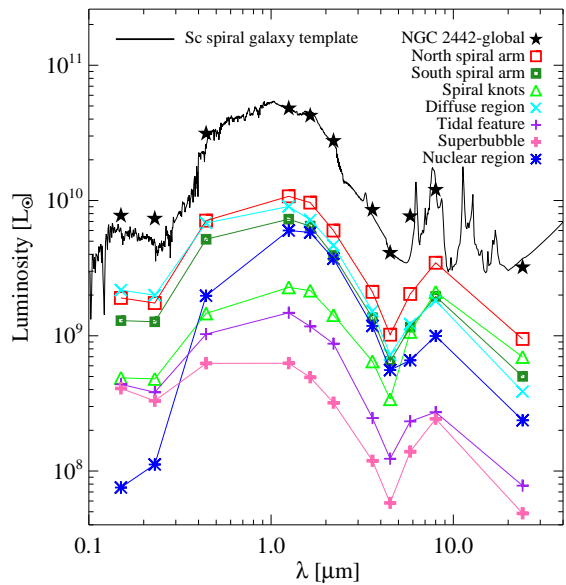}
\end{center}
\caption{SED of NGC 2442 showing the total luminosity in each band for the whole galaxy as well as individually for each region of interest. Template of an Sc spiral galaxy template from SWIRE library overlaid for comparison \citep{polletta07}. \label{fig_sed}}
\end{figure}

\section{Results}

\subsection{Spectral Energy Distribution and mid-IR colors \label{sect_sed}}
Figure\,\ref{fig_sed} shows the UV-to-IR spectral energy distribution (SED) of NGC 2442. For comparison, we overlay an Sc spiral galaxy template \citep{polletta07}, which is a good match to NGC 2442 in the near-IR to mid-IR, but is somewhat under-luminous in the UV.  Figure \ref{fig_sed} also shows the spectral energy distribution of each region of interest.  The SEDs of the northern and southern spiral arms are very similar although the northern arm is more luminous. Apart from those two masks, we find that, as expected, there is significant variation in the SED shape across the galaxy. The IR emission (at 24\um) appears strongly concentrated along the spiral arms (including the north and south spiral arms and the spiral knots).  By contrast, the UV emission is much more spread out with the strongest emission coming from the diffuse region.  We will return to this point in Section\,\ref{sect_sfr}. The nuclear region is extremely UV poor, and moreover has the reddest UV color suggesting high levels of obscuration (see Section\,\ref{sect_obsc}).  Relative to the rest of their SEDs, the strongest UV emitters are the tidal region at the tip of the southern spiral arm, and even more extremely the superbubble region. We discuss possible interpretations of the origin of the superbubble region in Section\,\ref{sect_superbubble}. 

\begin{figure}[h]
\begin{center}
\plotone{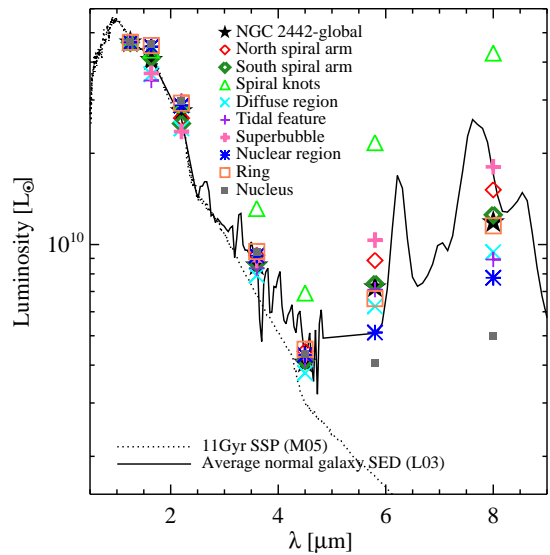}
\end{center}
\caption{Detailed near-IR SED of the total galaxy and regions of interest normalized to the J band.  An 11\,Gyr old SSP model \citep[M05;][]{maraston}, and a normal galaxies average spectrum \citep[L03;][]{lu03} are overplotted for comparison.  \label{fig_nir_sed}}
\end{figure}

To investigate more closely the relative strength of the polycyclic aromatic hydrocarbon (PAH) emission features that are believed to dominate the broadband 5.8\um\ and 8.0\um\ fluxes, Figure\,\ref{fig_nir_sed} shows a more detailed view of the near-IR SED for the total galaxy and regions of interest. The higher resolution of the J-8.0\,\umm bands allows for separation of the nuclear ring and nucleus, although the composite nuclear region is shown for comparison.  The SED for each region of interest has been normalized to the J band, representing a stellar mass normalization.  An 11\,Gyr old single stellar population (SSP) model from \citet{maraston} has also been plotted. The total galaxy SED is in very good agreement with the {\sl ISO} average spectrum of normal galaxies from \citet{lu03}. The observed excess over the pure stellar population model at 5.8 and 8.0\um\ is likely due to the 6.2 and 7.7\um\ PAH emission features.  \citet{lu03} suggest that the excess at $\sim$\,4\,--5\um\ seen in the spectra of normal galaxies is due to the presence of hot ($\sim$\,1000\,K) dust. We find evidence of this excess in all of our regions, suggesting hot dust is present throughout this galaxy.  We have tested this conclusion against a range of stellar population ages and both a burst and a nearly constant star-formation (e-folding time of 10gyr).  The spiral knots excess at 3.6\um\ is most likely due to the 3.3\um\ PAH emission feature.  

Compared to the small spreads in the H-4.5\,\umm bands (a product of our J-band normalization and the relatively constant old stellar populations spectral shape in this regime) for the different regions of interest, the spread in the 5.8 and 8.0\,\umm bands is 3-4 times larger, suggesting considerable variation in the relative PAH strength in the different regions. Relative to the stellar mass, the spiral knots and the superbubble region show the strongest PAH emission. In the nuclear region, the ring is the stronger infrared emitter, while the nucleus is more dominated by the stellar emission.

\begin{figure}[h]
\begin{center}
\plotone{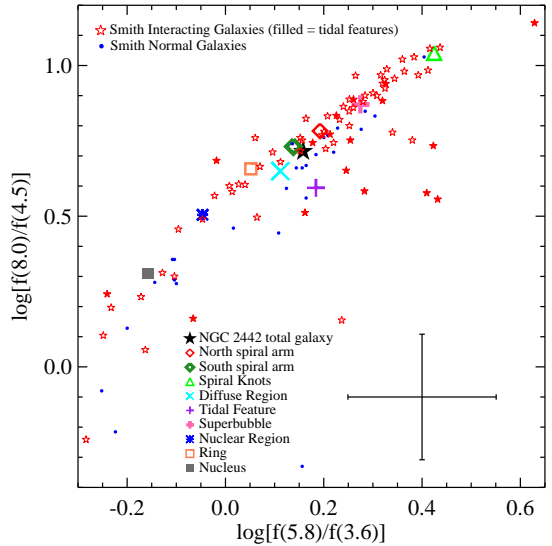}
\end{center}
\caption{IRAC color-color plot showing mean values for each region of interest and \citet{smith} data for normal and interacting galaxies.  A representative example of the spread in pixel values for each region of interest is shown by the large error bar in the bottom right.  \label{fig_irac}}
\end{figure}

Another common way to compare emission from the old stellar population, emitted in the 3.6 and 4.5$\mu$m bands, with emission from polycyclic aromatic hydrocarbons (PAH) in the 5.8 and 8.0$\mu$m, bands that trace young stellar populations, is the IRAC color-color plot.  \citet{sajina} find that star forming regions, old stellar populations, and active galactic nuclei are found in distinct locations on the plot.  Regions dominated by star formation are found at high color ratios while regions dominated by old stellar populations are found at lower color ratios.  Active galactic nucleus dominated regions are found at higher $f_{5.8}/f_{3.6}$ for a given $f_{8.0}/f_{4.5}$, due to a hot mid-IR continuum.  Figure \ref{fig_irac} shows the average colors of the regions of interest.  The colors are calculated using the total flux from each region of interest.  However, calculating the per-pixel color and then taking the average for each region gives us an estimate of the spread in color about the mean.  A representative 1$\sigma$ spread in per-pixel colors for the regions of interest is shown by the large error-bar.  The nuclear region average datapoint follows the linear trend of the other average datapoints and does not occupy the active galactic nucleus color-color space, suggesting that NGC 2442 does not have an IR active galactic nucleus, despite its low-ionization nuclear emission region (LINER) optical spectrum \citep{bajaja}.  The region that deviates the most from the general trend is the tidal debris region which lies to the right of the usual PAH/stars trend.  This could be due to: a) excess hot dust; b) varying 7.7/6.2 PAH emission ratio;  or c) a strong emission feature (s.a. molecular hydrogen) in the 5.8\um\ band. Without spectra, we cannot distinguish between these possibilities.

Figure \ref{fig_irac} also shows data from the interacting and normal galaxies samples studied by \citet{smith}, who also provide IRAC fluxes for the tidal features (e.g. bridges/tails) themselves.  Intriguingly, we find that the colors of the tidal features from \citet{smith} also tend to lie to the right of the usual relation, in agreement with our tidal region observations. 

\subsection{Superbubble \label{sect_superbubble}}
\begin{figure}[h]
\begin{center}
\plotone{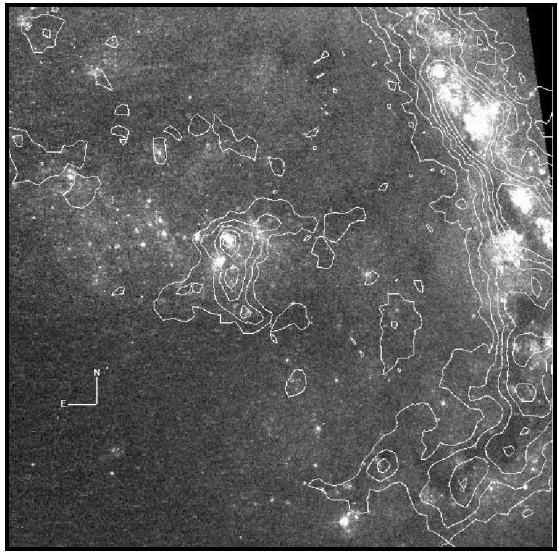}
\end{center}
\caption{The greyscale shows the H${\alpha}$ image for part of the galaxy, whereas the contours show the IRAC 8\um.  The image is centered in the superbubble region. \label{fig_ha_8um}}
\end{figure}

Figure\,\ref{fig_masks} shows a bubble-shaped region outside of the spiral arms in the southern part of the galaxy (centered at RA=7:36:16 and DEC=-69:33:12). The diameter of this `superbubble' is $\sim$\,1.7\,kpc.  Comparable sized gas superbubbles have previously been observed in starburst galaxies \citep[e.g.][]{vader95,tsai09}, and are generally believed to be the result of hundreds of supernovae whose individual supernova remnants (SNR) merge to form such a structure. In Figure\,\ref{fig_ha_8um}, we show a comparison (over a wider region) of the ACS F658N H${\alpha}$ image (available from the MAST archive, PI S. Smartt) and the IRAC 8\um\ contours. The first conclusion we can draw is that there is significant H${\alpha}$ emission in this region (this area is also a significant UV emitter, see Figure\,7). The H${\alpha}$ distribution in this image is not bubble-like (which is not surprising given the likely age of this structure).  The H${\alpha}$ emission is in good agreement with our result  that the superbubble region has among the highest star-formation intensities in NGC2442 (see Table\,3).  This area has significant radio emission (including polarized emission) as can be seen in the 6cm radio data presented in \citet{harnett}. The recent supernova SN1999ga \citep{pastorello} is located right at the edge of this superbubble. Further observations are needed to more fully understand the nature of this region; however, our current data suggests a large star-forming region outside the spiral arms, whose morphology is likely to have been shaped by generations of supernovae. 

\subsection{Dust Obscuration \label{sect_obsc}}
The UV obscuration, $\rm{A_{1500}}$, and the UV slope, $\beta$, defined as $f_{\lambda}\propto\lambda^{\beta}$, have been shown to be correlated, especially for starburst galaxies \citep{meurer}. In terms of the {\sl GALEX} FUV and NUV fluxes, $\beta$ is derived by:
\begin{equation}
\label{eq_beta_uv}
\beta=\frac{\log(f_{\nu,1500}/f_{\nu,2300})}{\log(1500/2300)}
\end{equation}
where $f_{\nu,1500}$ is the specific flux at 1500\AA.  To understand how the observed UV slope is related to the UV obscuration, we can re-arrange this to show that:
\begin{equation}
\label{eq_beta_gen}
\beta=\beta_o+2.155A_{1500}[1-(\kappa_{2300}/\kappa_{1500})],
\end{equation}
where $\beta_o$ is the dust-free UV slope and the ratio  $\kappa_{2300}/\kappa_{1500}$ incorporates the UV slope of the extinction curve. The most widely used relation is that of \citet{meurer}, who find that for UV-selected starburst galaxies $\beta$ and $\rm{A_{1600}}$ are related by:
\begin{equation}
\label{eq_a_beta}
\rm{A_{1600}}=4.43+1.99(\beta-2.0) 
\end{equation}
Note that this differs slightly from the form of the \citet{meurer} relation because their $\beta$ is defined in terms of $f_{\lambda}$ and ours in terms of $f_{\nu}$. The uncertainty in $\rm{A_{1600}}$ is 0.08 mags and in $\beta$ it is 0.04.  We take A$_{1600}$\,=\,A$_{1500}$ (since this correction is expected to be relatively negligible). 
We begin by assuming this relation; however, we return to the question of the most appropriate extinction curve in Section\,4.5. 

Based on Equation\,\ref{eq_a_beta}, the area of highest obscuration is the nuclear region with an average obscuration of 4.4\,mags, while the rest of the galaxy is obscured by $\sim$1.4-2.4\,mags.  The average obscuration at 1500\AA{} is given in Table \ref{table_masks2} for each region of interest.

The major limitation with the \citet{meurer} UV-slope relation is that it is derived using starburst galaxies and shown to be a good approximation when the UV emission consists of ionizing UV photons from young stars.  \citet{calzetti} find that only 40\% of the UV emission from the normal galaxy M51 traces current star formation, defined as populations less than 30\,Myr old.  This suggests that the UV-slope of normal galaxies will be affected by the older stellar population instead of only representing the preferential extinction of UV light from young stars for which it was derived.

In addition, \citet{salim07} calibrate relations between {\sl GALEX} FUV and NUV bands and ${\rm A_{1500}}$ for star-forming galaxies.  The relation for galaxies with ${\rm (FUV-NUV)< 0.95}$, as found in NGC2442, is given by:
\begin{equation}
{\rm A_{1500}=3.32(FUV-NUV)+0.22}
\end{equation}
We expect this relation to give a better estimate of the dust extinction in NGC2442 than the Meurer relation as it is calibrated on normal star-forming galaxies rather than starbursts.  This is discussed in more detail in Section\,4.5.

\begin{figure*}[h]
\begin{center}
\plottwo{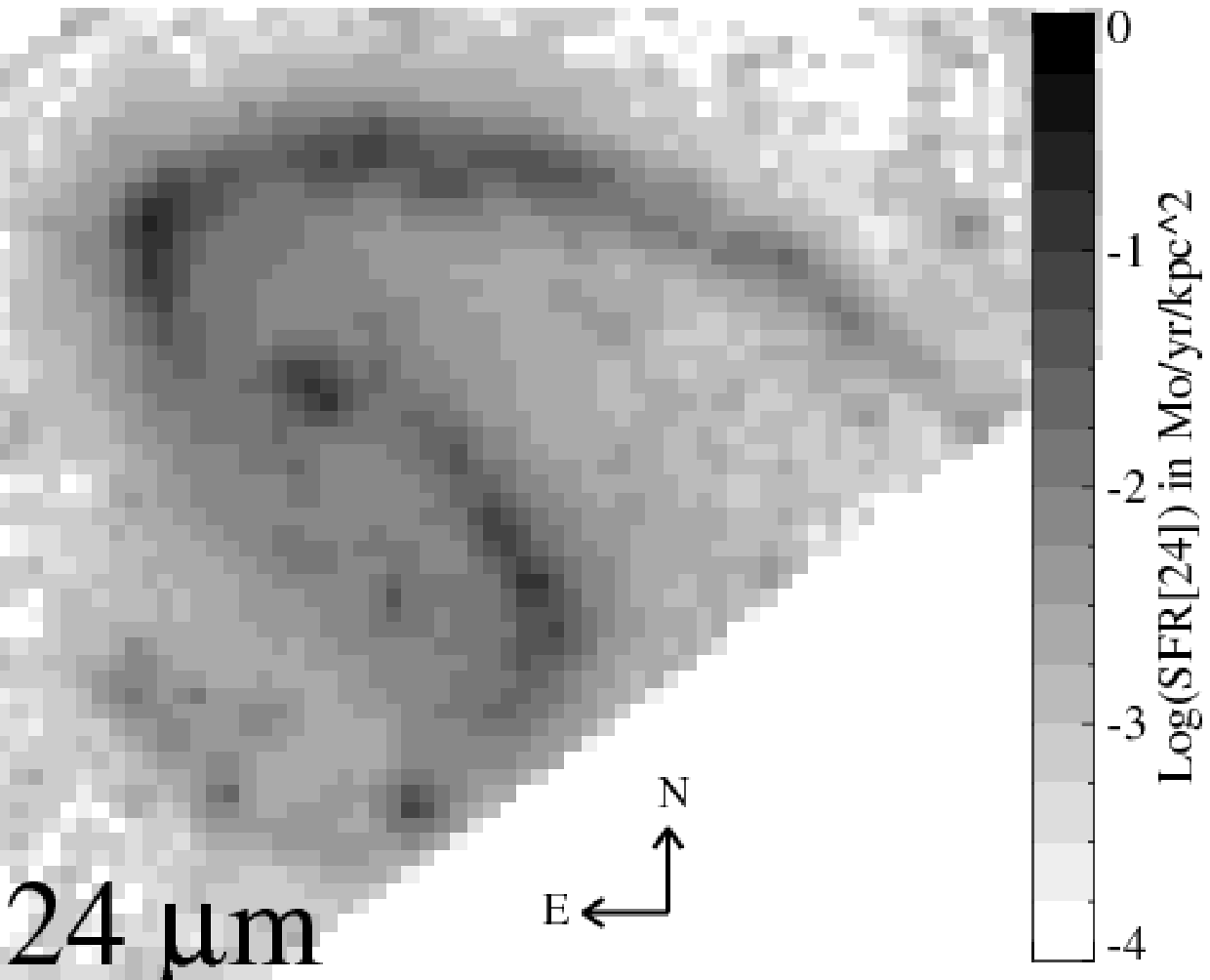}{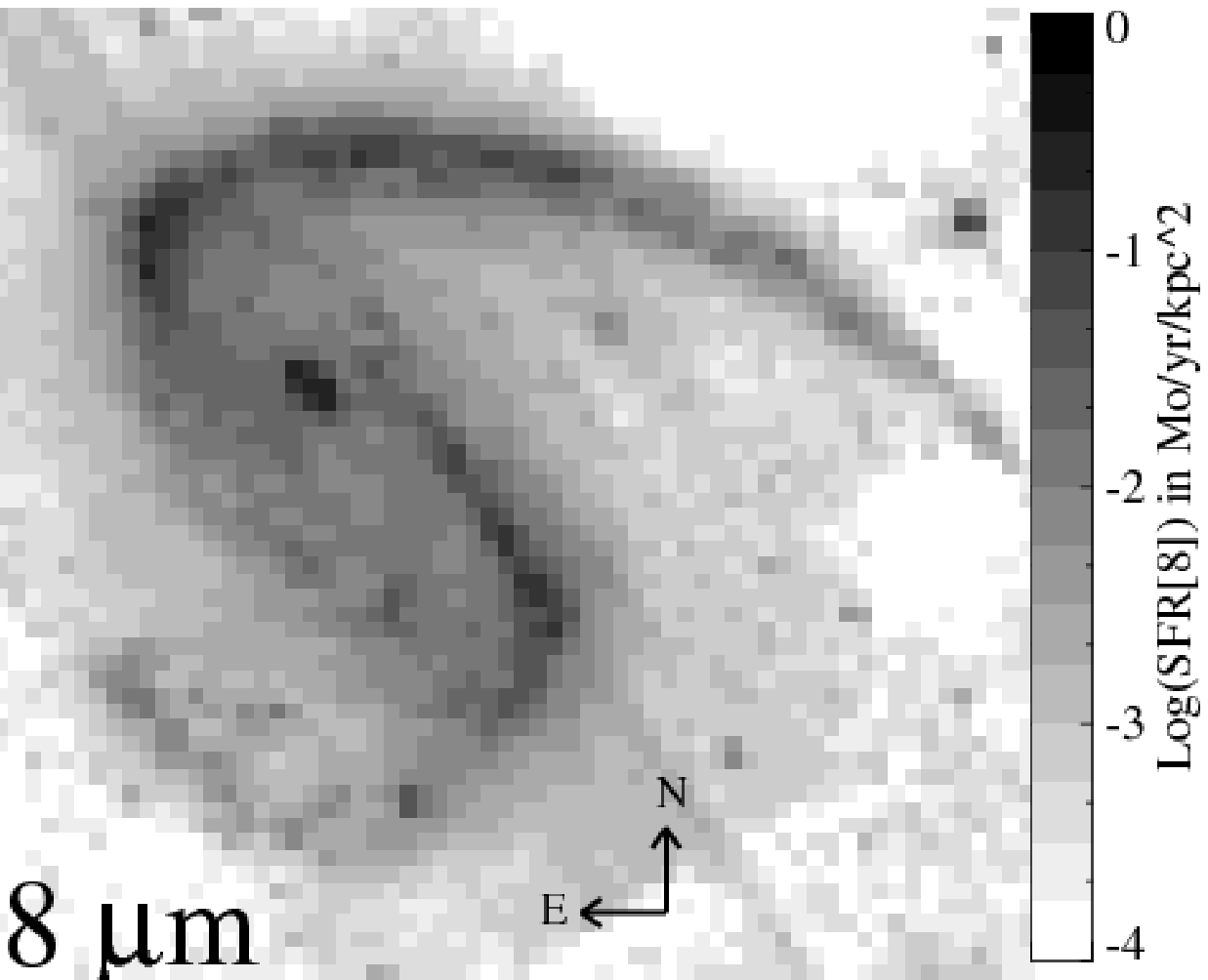} 
\plottwo{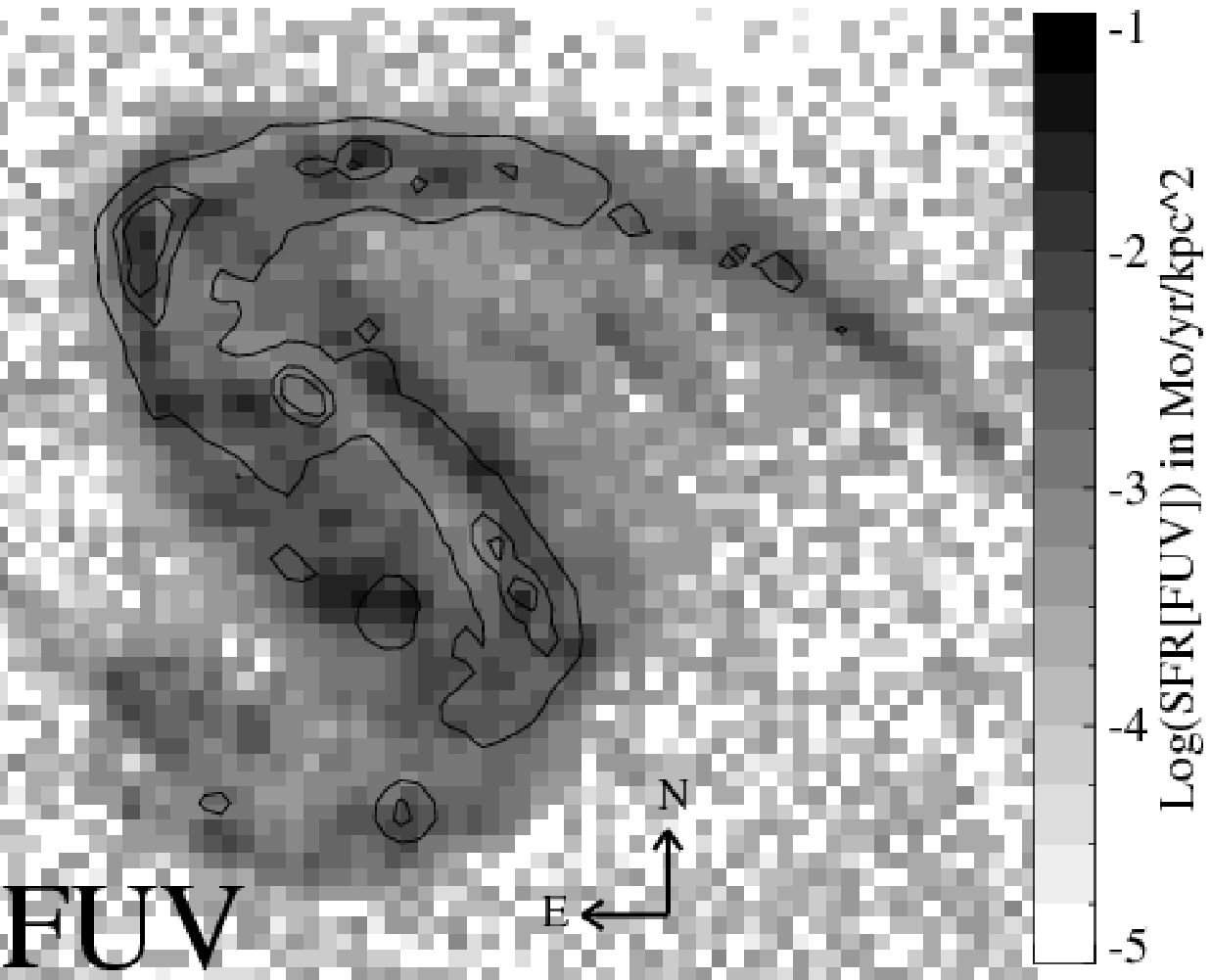}{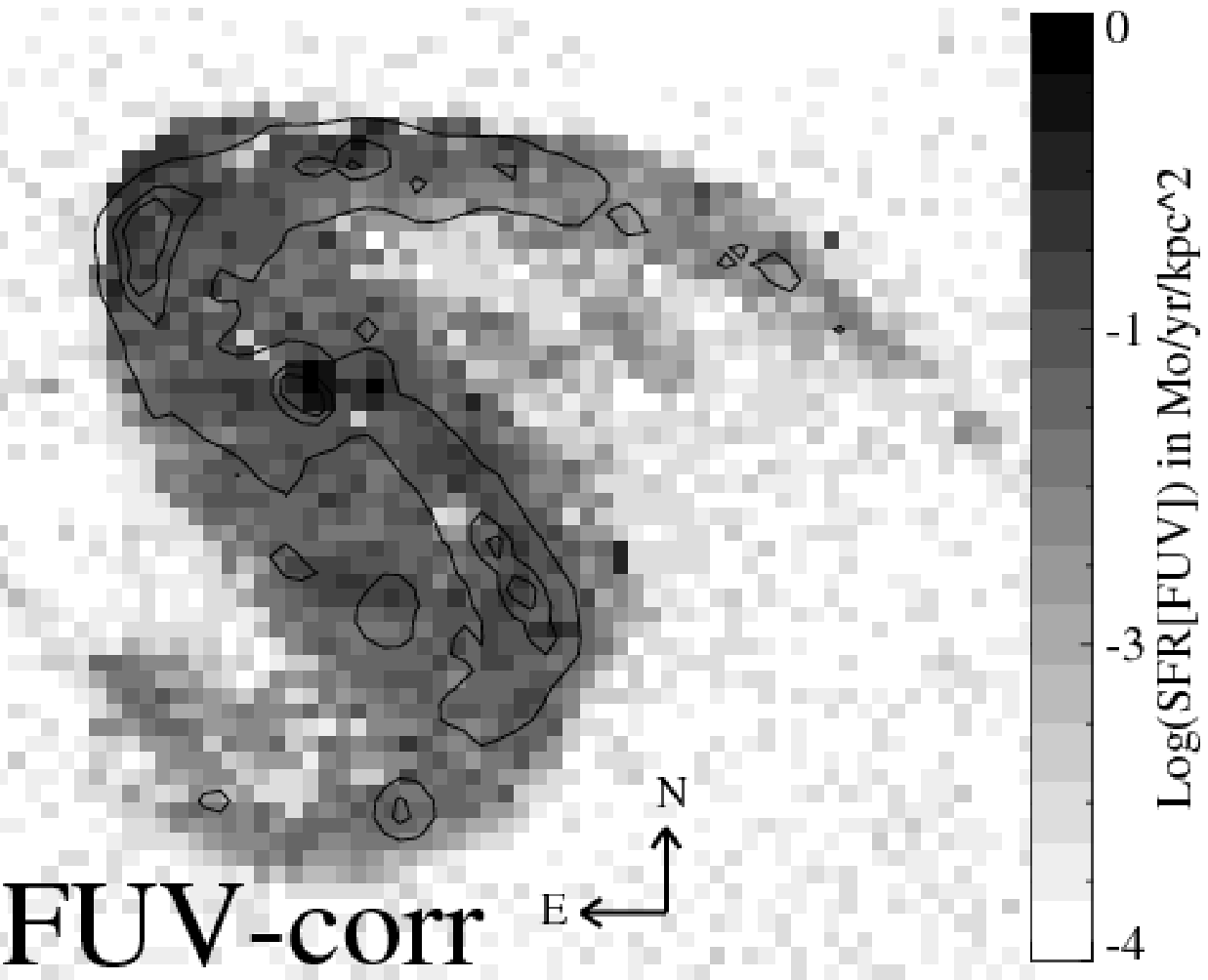}
\plottwo{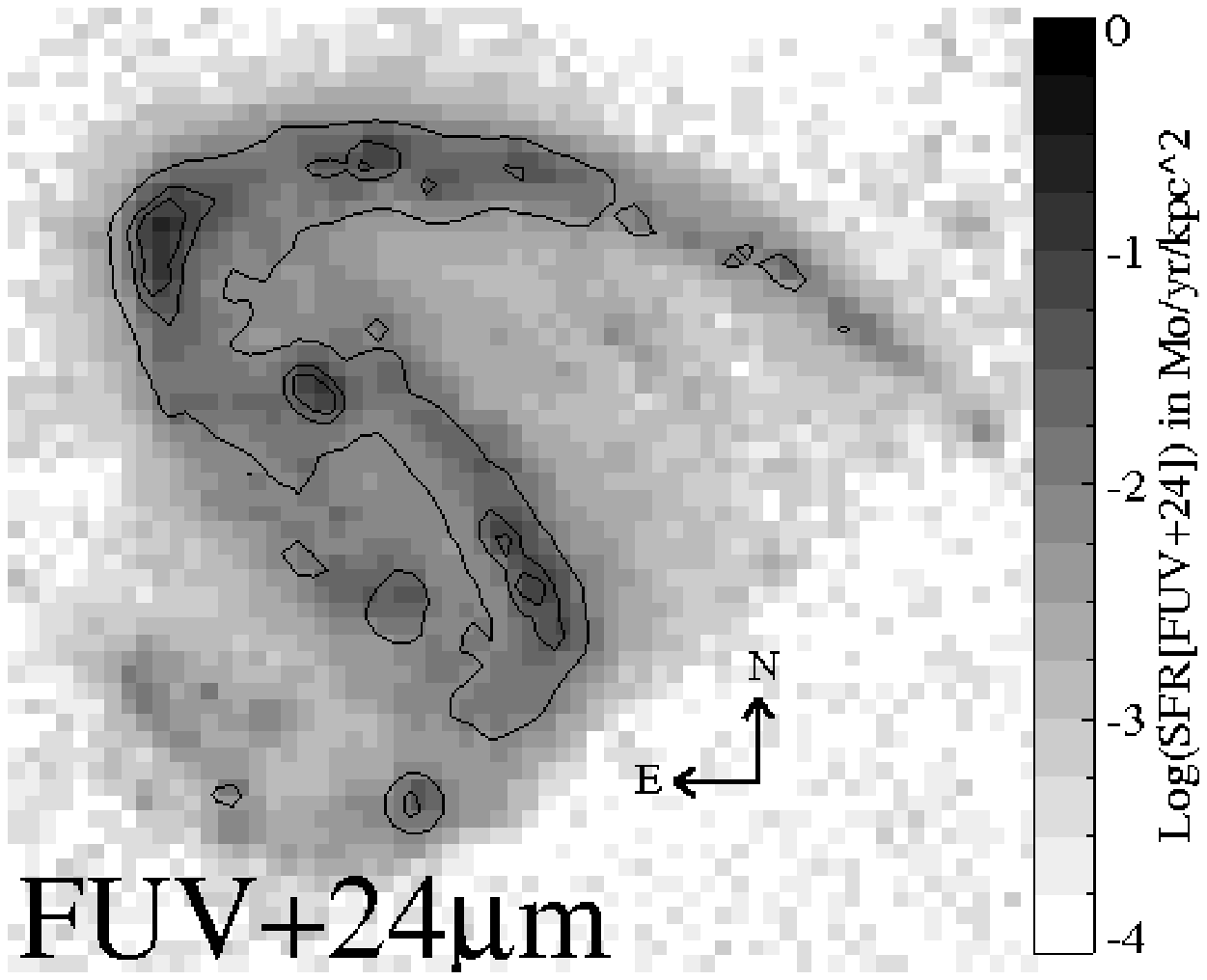}{fig6f.eps}
\caption{SFR maps from 24\,$\mu$m, 8\,\um, as well as dust-uncorrected and dust-corrected FUV emission and FUV+24\,\um.  The SFR(FUV) and SFR(FUV+24) maps have SFR(24) diagram contours overplotted at values of 0.01, 0.05 and 0.1\,$\rm{M_{\sun}yr^{-1}kpc^{-2}}$. Note the different scales of SFR intensity in the case of dust-uncorrected SFR(FUV).  \label{fig_sfr}}
\end{center}
\end{figure*}

\subsection{Star Formation Rate\label{sect_sfr}}

\subsubsection{Comparing SFR Indicators and the SFR maps of NGC2442 }

In the literature, there is a wide array of relations to convert the FUV, 8\um, and 24\um luminosities to SFR. Since each SFR relation is calibrated using a different subset of galaxies or theoretical models, it is important to consider their applicability to NGC2442.  As NGC2442 appears to be essentially an $L^*$ galaxy (see Section\,\ref{sec_global}), we choose to apply SFR relations that are calibrated using normal star-forming galaxies.  For greater internal consistency, and to allow direct comparison with the overall SFR(H$\alpha$) presented in Section\,5, we also choose infrared conversion relations that are all ultimately calibrated on the SFR(H$\alpha$) relation \citep{kennicutt98}. Because we are looking at resolved star-formation in NGC2442 we consider both relations calibrated on a local and a global scale. 

For a UV-based SFR, we consider both the relations in \citet{kennicutt98} and in \citet{salim07}. The latter relation gives SFR(FUV) values $\sim$30\% smaller than the former.  \citet{kennicutt98} assume a \citet{salpeter} initial mass function (IMF) and a constant star-formation. The effect of recent starbursts would paradoxically be to lower the derived SFR values (as higher UV/SFR).   \citet{salim07} derive the SFRs of nearby galaxies by fitting their UV+optical SEDs and averaging their star-formation over the past 100Myrs.  These are used to calibrate a conversion relation between $L_{\rm{FUV}}$ and SFR. This relation generally results in SFRs that are $\sim$\,30\% lower than those based on \citet{kennicutt98} when it is converted from a Chabrier \citep{chabrier} to Salpeter IMF. This is due, in part, to a departure from a constant star-formation, which, as discussed above, tends to result in lower SFR.  The difference between these two relations is within the large systematic uncertainty quoted in \citet{kennicutt98}; however, since NGC2442 has a tidally distorted morphology and regions of higher SFR intensity (as is discussed later in this Section), it is more likely that there were recent bursts rather than a constant star-formation history over the past $\ls$\,100\,Myr. Therefore, we adopt the \citet{salim07} relation. We adopt a 0.15\,dex 1\,$\sigma$ systematic uncertainty on SFR(FUV) \citep[see][]{kennicutt98,salim07}. 

We derive SFR(8) and SFR(24)$_{global}$ based on the relations presented in \citet{zhu08}. These are calibrated based on the correlation between total integrated mid-IR  luminosities of normal galaxies with the extinction corrected H$\alpha$ luminosities, and applying the \citet{kennicutt98} H$\alpha$-SFR relation.  However, because we are looking both for the global and local SFRs, we also look at SFR relations calibrated on a local scale. Specifically, we use the SFR-$L_{24}$ relation based on individual H{\sc ii} regions of normal star-forming galaxies given in  \citet{relano07} (${\rm SFR(24)_{HII}}$). This again is calibrated using the extinction corrected  H$\alpha$ luminosities and the \citet{kennicutt98} relation. We also find this to be consistent with the \citet{calzetti} $L_{24}$\,-\,$L_{P\alpha}$ relation for HII regions in M51, if the SFR(P$\alpha$) relation from \citet{alonso} is adopted .  For more discussion on different SFR indicators and their associated uncertainties see \citet{kennicutt98} and \citet{calzetti10}.  Lastly, we also use the  composite relation, SFR(FUV+24), derived by \citet{bigiel08}. Their relation aims at deriving the star-formation intensity both on a global and local scale.  It is empirically derived, based on both the SFR(H$\alpha$+24) relation in \citet{calzetti07} and the SFR(FUV) relation in \citet{salim07}. Below, we summarize the relations we use. 

\begin{align}
& \rm{SFR}_{FUV}[\rm{M_{\odot}} yr^{-1}]=2.09\times10^{-10}(L_{1600A}\rm[L_{\odot}]) \label{eq_sfr_fuv}\\
& \rm{SFR}_{8\mu m}[\rm{M_{\odot}} yr^{-1}]=6.33\times10^{-10}(L_{8\mu m}\rm[L_{\odot}]) \label{eq_sfr_8}\\
& \rm{SFR}_{24\mu m,\,global}[\rm{M_{\odot}} yr^{-1}]=1.42\times10^{-8}(L_{24\mu m}\rm[L_{\odot}])^{0.848} \label{eq_sfr2_24} \\
& \rm{SFR}_{24\mu m,\,HII}[\rm{M_{\odot}} yr^{-1}]=3.44\times10^{-9}(L_{24\mu m}\rm[L_{\odot}])^{0.97} \label{eq_sfr_24}
\end{align}

\begin{figure}[h]
\begin{center}
\plotone{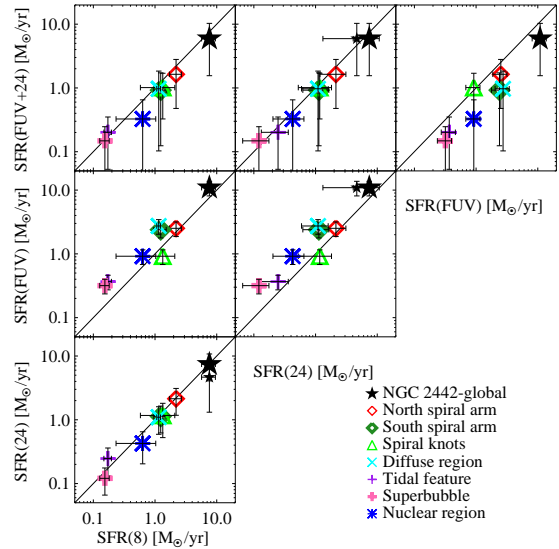}
\end{center}
\caption{Comparison of SFR tracers, including SFR(24), SFR(8), and SFR(FUV), where the FUV has been corrected for dust obscuration using the UV-slope.  The error-bars show 1$\sigma$ uncertainty. The SFR(24) used is the SFR(24)$_{\rm{HII}}$ one. SFR(24)$_{global}$ is shown only for the total galaxy (the smaller black star symbol).  \label{fig_sfr_compare}}
\end{figure}

Figure\,\ref{fig_sfr} shows the SFR intensity maps based on these relations. The areas of highest SFR intensity in decreasing order are the spiral knots, the nuclear region, and the spiral arms.  Our SFR and SFR intensity estimates both globally and per region are presented in (see Table\,\ref{table_masks3}).   We compare the dust corrected-SFR(FUV), SFR(8), ${\rm SFR(24)_{HII}}$, and SFR(FUV+24) values for the total galaxy and regions of interest in Figure \ref{fig_sfr_compare}, where ${\rm SFR(24)_{global}}$ has been calculated for only the total galaxy region and plotted with the smaller symbol. 

Only the North spiral arm and the spiral knots show agreement between all SFR indicators within 1\,$\sigma$. In particular, this suggests that the starburst-calibrated dust correction in the FUV is applicable for these regions (we return to this point in  Section\,4.5).  A significant outlier here is the superbubble region, which we speculate is due to the presence of a very recent starburst in that region. As expected, the SFR(8) and SFR(24) show the best agreement. Although it is surprising that this good agreement (even for the total galaxy) is not for the SFR(8) and SFR(24)$_{global}$ values which are calibrated on the same sample and in the same method \citep{zhu08}. This may indicate that the influence of H{\sc ii} regions in our galaxy is somewhat stronger than average in the \citet{zhu08} sample. The one somewhat outlying point is for the tidal region -- in the sense that it is deficient in 8\um\ emission compared to its 24\um\ emission relative to the average for the galaxy. We already discussed that this region has somewhat anomalous PAH ratios in Section\,4.1.  The tidal region and the superbubble region are the only ones where the SFR(8) and SFR(24)$_{global}$ values are closer than  SFR(8) and SFR(24)$_{\rm{HII}}$.

It is important to note that the error bars shown in Figure\,\ref{fig_sfr_compare} for the SFR(FUV) show the uncertainty in the dust correction; as apparent in Table\,\ref{table_masks3}, the uncertainty in the overall SFR(FUV) relation is much larger.  Correcting the FUV emission for dust extinction introduces larger uncertainties for SFR(FUV) than for SFR(8) and SFR(24) that are only surpassed by the large intrinsic spread of 0.3\,dex in SFR(FUV) relations described by \citet{kennicutt98}.  A possible systematic uncertainty comes from using a dust obscuration correction calibrated for starburst galaxies, such as the UV-slope, which would overestimate the dustiness and L(FUV) for a normal star-forming galaxy \citep[for a discussion of this problem see][]{boquien209}.  For the uncertainty in SFR(8), we emply two additional SFR(8) relations, those of \citet{calzetti} and \citet{wu05}, and use half the spread in the three relations for each region of interest to define the uncertainty.  We use the same method to estimate the uncertainty in SFR(24), but employing a larger range of relations, including those of \citet{alonso}, \citet{calzetti}, and \citet{calzetti07},  a linear SFR relation given in \citet{calzetti10} and derived in \citet{zhu08}\footnote{\citet{zhu08} find that the data is better fit by a non-linear model (the relation we adopt), but also derive a linear SFR relation.}, and that of \citet{rieke09}.  The relations given by \citet{alonso} and \citet{calzetti07} are calibrated using starbursts, however we do not find a very large discrepancy between these different SFR(24) indicators.  The largest discrepancy is actually introduced by the linear relation in \citet{zhu08}.

\begin{figure}[h]
\begin{center}
\plotone{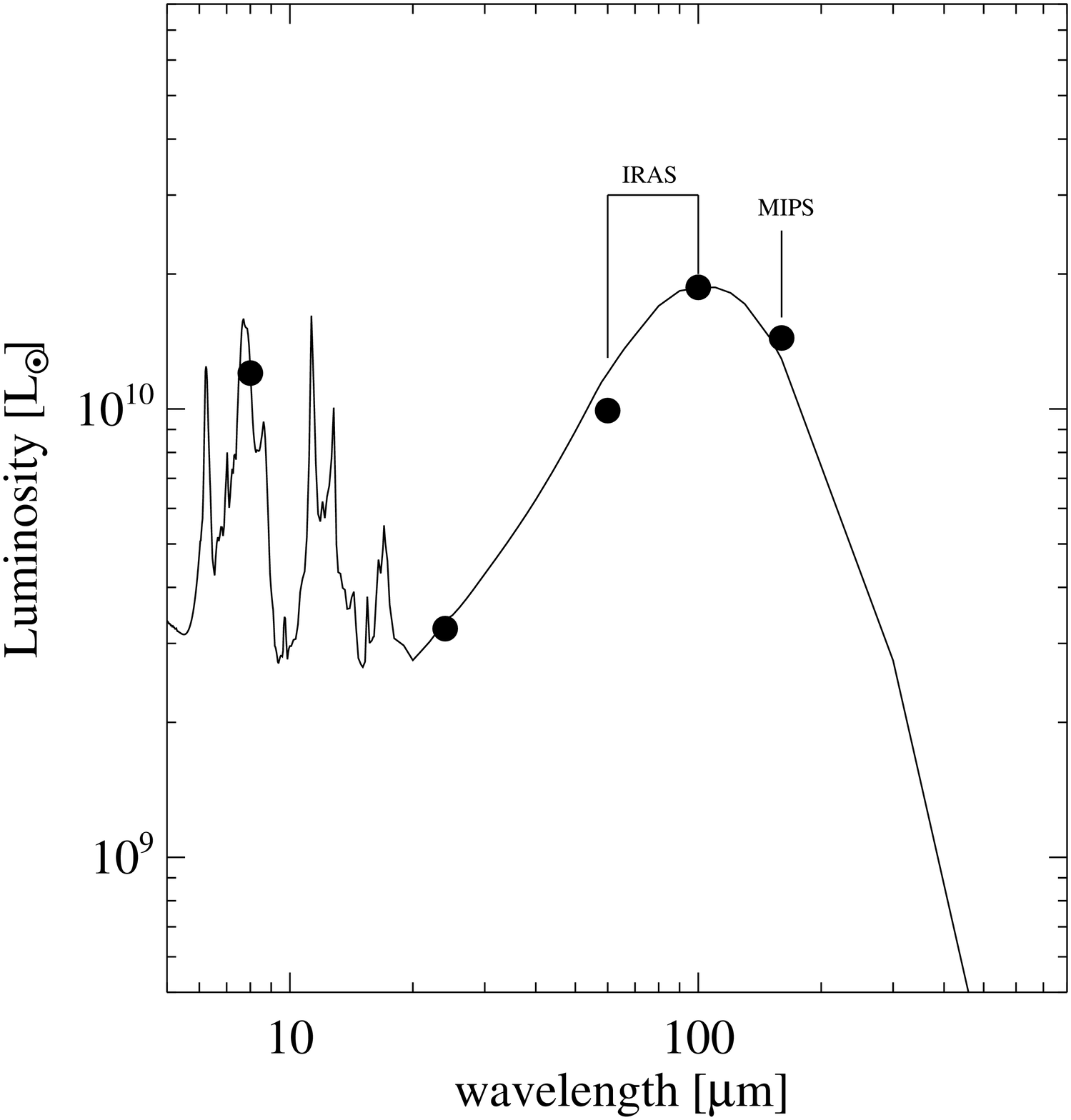}
\end{center}
\caption{The global far-IR SED of NGC2442. The SED template is the same as shown in Figure\,2 (i.e. an Sc galaxy template). Along with the total galaxy 8\um\ and 24\um\ points as given in Table\,1, we also show the IRAS\,60 and 100\um\ points (from NED) and the MIPS\,160\um\ point (we measured 57.5Jy for the total galaxy) \label{fig_fir}}
\end{figure}

\subsubsection{Global SFR indicators from H$\alpha$ and far-IR emission \label{sec_global}}
\label{sect_sfrfir}
For the ease of subsequent analysis we also derive the global SFR for NGC2442 based on two well established SFR indicators: H$\alpha$ and the total infrared emission, $L_{IR}$.  The only previously published estimate of the SFR in NGC 2442 was based on its H$\alpha$ emission \citep{mihos}.  Updating their value to account for the different cosmology we adopt as well as  to the SFR relation in  \citet{kennicutt98}, we find SFR(H$\alpha$)=4.2\,$\rm{M_{\sun}yr^{-1}}$ for the whole galaxy. NGC2442 has constraints at longer wavelength (far-IR emission) based on IRAS observations (at 60 and 100\um) and MIPS 160\um\ observations. In order to accurately estimate the total $L_{\rm{IR}}$, we can use the contraints at longer wavelengths to choose a generic galaxy template, such as for an Sc galaxy, that fits the data and integrate the luminosity in a defined range.  We calculate the integrated $L_{\rm{FIR}}$\,$\equiv$\,$L_{8-1000}$, giving us $\log(L_{\rm{FIR}}/L_{\odot})$\,=\,10.53 and SFR(FIR)\,=\,5.85\,\msun/yr, based on \citet{kennicutt98}. Recently, \citet{kennicutt09} derived global composite $H_{\alpha}$ and infrared SFR relations for the SINGS galaxies., which as discussed before are similar to NGC2442 in luminosity and hence these relations are applicable in our case.  \citet{kennicutt09} derive relations based both on the 24\um\ emission and total IR (where TIR is defined as $L_{3-1000}$). Based on these SINGS-calibrated relations, we obtain SFR(H$_{\alpha}$+24)\,=\,6.1\msun/yr  and SFR(H$_{\alpha}$+TIR)\,=\,7.1\msun/yr. These are very similar to our total galaxy SFR(FUV+24) and  SFR(24)$_{\rm{HII}}$ respectively.  Combined,  these relations account for both direct/unobscured young stellar light, the H{\sc ii} regions associated with ongoing SF and the total dust-reprocessed largely young stellar light. This leads us to believe that the true SFR is likely in the range $\sim$\,6\,--\,7\msun/yr.  

\begin{figure}[h]
\begin{center}
\plotone{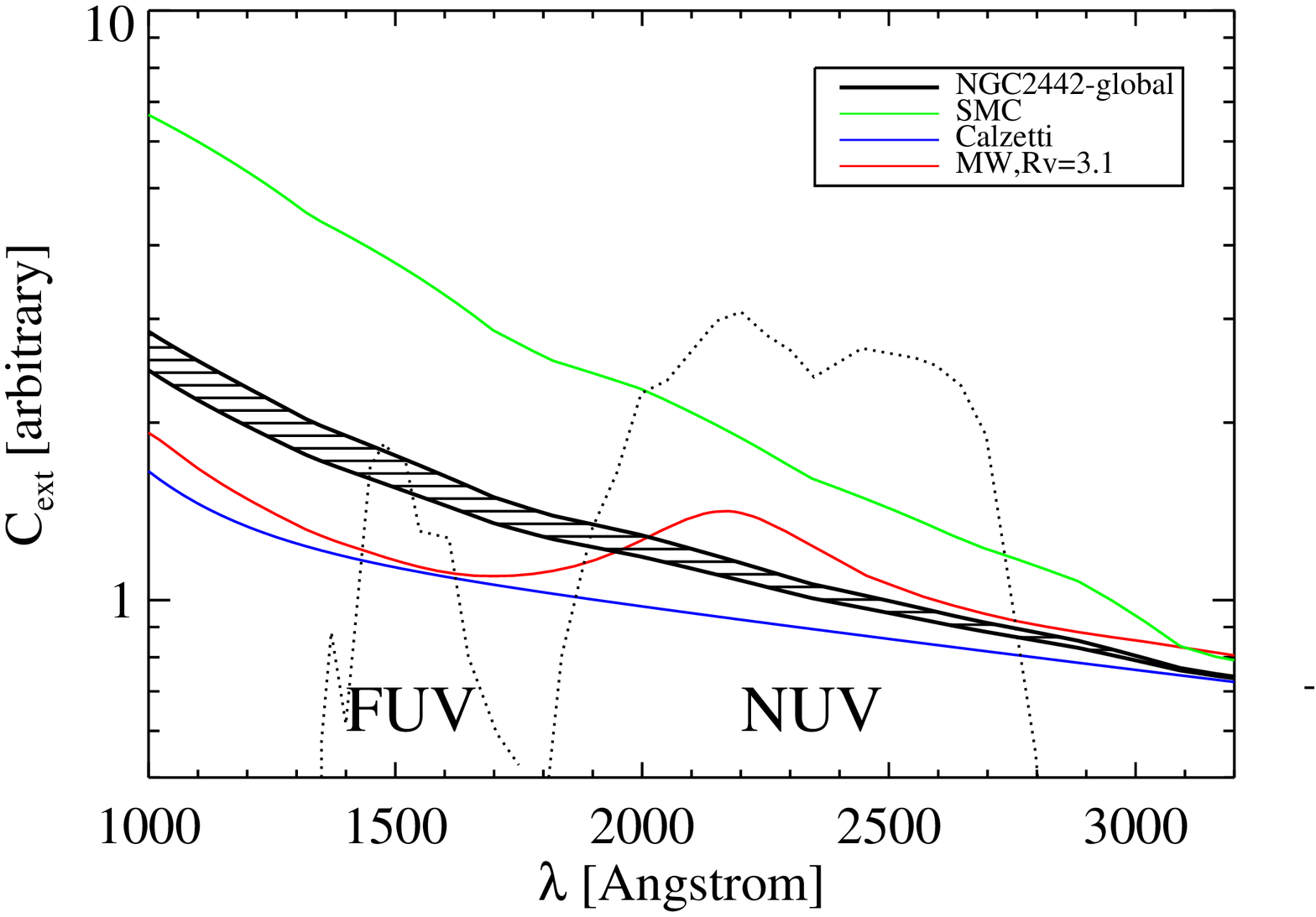}
\end{center}
\caption{The extinction curve we need to reconcile the UV and IR based SFR estimates. Our estimates are found to lie between the SMC and Calzetti extinction laws and are defined as linear combinations of the two. The hashed region represent the range of 6\,--\,7\msun/yr where the true SFR for NGC2442 is most likely to lie. 
\label{fig_ext}
}
\end{figure}

\subsection{Constraints on the extinction curve in NGC2442 \label{sec_extinction}}

In the previous section, we showed that using the \citet{meurer} relation to translate the observed UV slope to the UV extinction leads to larger SFRs than obtained from the IR-based relations. This is assuming the \citet{salim07} relation for SFR(FUV) and the discrepancy is even larger if the \citet{kennicutt98} relation is used (which gives SFR(FUV) that are $\sim$\,30\% larger).  As discussed previously, the systematic uncertainty in SFR(UV) is quite significant, but for the purposes of this section it is ignored.  

In Section\,\ref{sec_global}, we argued that the true SFR is likely to lie in the range $\sim$\,6\,--\,7\msun/yr. Without dust correction our SFR(FUV) for the total galaxy is 1.7 implying $A_{FUV}$\,=\,1.4\,--\,1.5. The \citet{meurer} relation gives us $A_{FUV}$\,=\,2.0.  We use the generic $A_{1600}$ vs. $\beta$ relation given in Equation\,\ref{eq_beta_gen}, to constrain the UV-slope of the extinction curve that would reconcile the UV and IR-based SFR estimates.  \citet{meurer} already discuss that, for star-forming galaxies, the range in the dust-free UV slope, $\beta_o$, is relatively small (-2.6 to -2.0) and adopt the value of $\beta_o$\,=\,-2.23. We adopt the same intrinsic slope. In order to derive the $\kappa_{2300}/\kappa_{1500}$ ratio for different extinction curves, we convolve any given curve with the {\sl GALEX} FUV and NUV filters. We find that, for a typical MW extinction curve,  the additional absorption in the NUV band due to the 2200\AA\ absorption feature leads to even larger dust-corrected UV-based SFR estimates, making such relatively `flat' extinction curves unlikely.  We then concentrate on `feature-free' and relatively steep extinction curves. The best known such extinction law is the one observed in the SMC \citep{prevot84}\footnote{Here we made use of: \url{http$://$webast.ast.obs$-$mip.fr$/$hyperz$/$hyperz$\_$manual1$/$node10.html}}. In Figure\,\ref{fig_ext}, we show the SMC extinction law, the Calzetti starburst extinction law \citep{calzetti01} and the MW $R_V$\,=\,3.1 extinction law \citep{draine03} compared with the {\sl GALEX} filters.  We find that reconciling the global 24\um\ SFR with the UV SFR requires an extinction law that lies in between the two curves.  This can be described as a simple linear combination: C$_{\rm{ext}}$\,=\,0.81\,C$_{\rm{ext,Calzetti}}$+0.19\,C$_{\rm{ext,SMC}}$ or C$_{\rm{ext}}$\,=\,0.61\,C$_{\rm{ext,MW}}$+0.39\,C$_{\rm{ext,SMC}}$, where the latter largely reflects the influence of the 2000\AA\ feature . We can also compare this with the modified $A_{FUV}$ relations found by \citet{salim07}, who find that for their large sample of local UV-detected galaxies, the \citet{meurer} relation (being specifically geared toward starbusts) generally overestimates $A_{FUV}$. Using the \citet{salim07} relation, we find $A_{FUV}$\,=\,1.42, which would make the dust-corrected SFR(FUV) much more consistent with our best estimate of the true SFR.  

We repeat this exercise on a region-by-region basis to potentially look for changes in the extinction curve across the galaxy. Table\,\ref{table_masks2} shows the values for $A_{1600}$ we derived following both \citet{meurer} and \citet{salim07}, as well  the $A_{1600}$ required to match the SFR(24)$_{\rm{HII}}$. We chose SFR(24)$_{\rm{HII}}$ as it gives the largest values (compared with SFR(24)$_{global}$ and SFR(FUV+24)) and therefore in most cases is the closest to SFR(FUV)$_{\rm{corr}}$. We discuss different choices below.  We find that, as for the total galaxy, the \citet{salim07} relation for $A_{1600}$ leads to closer agreement between SFR(FUV) and SFR(24) for most regions. Considering the typical scatter of the $A_{1600}$ relation in \citet{salim07}, we consider the match good for both spiral arms, as well as for the tidal and nuclear regions.  Therefore their extinction curves are probably not too dissimilar to that shown in Figure\,\ref{fig_ext}. The cases where there is significant deviation are the diffuse region, superbubble and spiral knots. In both the case of the diffuse region and the superbubble, the UV slope suggests far more reddening than needed to reconcile SFR(FUV) and SFR(24). In this case, choosing SFR(24)$_{global}$ leads to even worse agreement, and choosing SFR(FUV+24) makes little difference.  Considering the full spread of the SFR(FUV) uncertainty (which is dominated by the systematic uncertainty on the IMF and star-formation history), if we go to the upper limits of SFR(FUV) for these two regions, the discrepancy will be even larger. However the lower limits on SFR(FUV) could bring the dust corrected SFR(FUV) in agreement with SFR(24). As discussed previously, this may arise as a result of a significant, recent starburst (as can be seen in the H$\alpha$  emission in the superbubble).  A steeper extinction curve would also help reconcile the two  SFR estimates. However,  we find that for example, to obtain the $A_{\rm{1600,IR-matched}}$ for the superbubble, we would need an unrealistic extinction curve that is much steeper than the SMC. The influence of recent starburst activity is much more likely. On the other hand the spiral knots, are the only region where the \citet{meurer} relation for $A_{1600}$ leads to better agreement between SFR(FUV) and SFR(24).  Therefore, an extinction curve closer to the \citet{calzetti01} starburst extinction is very likely. This is also not surprising given that this region has the strongest SFR intensity in this galaxy.  The two are brought into even better alignment if the SFR(FUV+24) is considered instead of SFR(24)$_{\rm{HII}}$.

\begin{figure}[h]
\begin{center}
\plotone{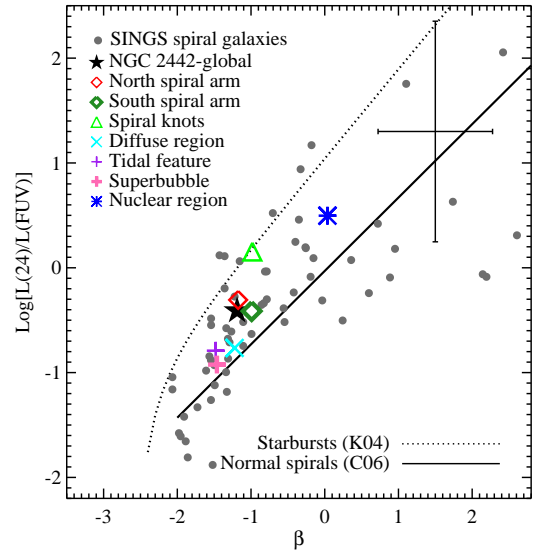}
\end{center}
\caption{A modified IRX-$\beta$ plot showing the average values for each region of interest as well as the total galaxy.  The large errorbar shows the typical spread in per pixel values for each region. We also show relations for starbursts from \citet[K04;][]{kong04} and for normal spirals from \citet[C06;][]{cortese06}.  The gray filled circles shows SINGS spiral galaxy datapoints from \citet{dale07}. \label{fig_meurer}}
\end{figure}

\subsection{The IRX-$\beta$ Relation \label{sect_meurer}}
The IRX-$\beta$ relation compares L(TIR)/L(FUV) and $\beta$, but since we do not have the total IR luminosity for the regions of interest in NGC 2442, we use L(24)/L(FUV) and $\beta$.  This modified IRX-$\beta$ diagram is shown in Figure \ref{fig_meurer},  showing the average for each region of interest, as well as for the total galaxy.  We find slightly different averages for the regions of interest when we calculate L(24)/L(FUV) and $\beta$ using the total flux for each region versus when we calculate the per-pixel values and average them for each region.  The representative error-bar shown in Figure\,\ref{fig_meurer} is the 1$\sigma$ spread in per-pixel values around the mean, not the error in the mean, which is much smaller given the large number of pixels in most regions.  This plot also show the relation for starbursts from \citet{kong04} and that for normal spirals from \citet{cortese06}.  To convert between L(TIR) and L(24) we use the method of Section \ref{sect_sfrfir} to compute the ratio of L(TIR)/L(24).  The L(TIR) of \citet{kong04} is defined between 3-850\umm and we estimate L(TIR)/L(24)$\sim$11.37.  The L(TIR) of \citet{cortese06} is defined between 1-1000\umm and we estimate a larger ratio of L(TIR)/L(24)$\sim$21.39.  The data for SINGS spiral galaxies from \citet{dale07} have been included to show the range of L(24)/L(FUV) versus $\beta$ values for a normal star-forming population. We find that our spiral knots alone are in good agreement with the starburst relation, while the rest of the regions and the total galaxy as a whole are in good agreement with the normal galaxies relation. \citet{boquien209} argue that this spread between the normal star-forming galaxies and starbursts can be explained by differences in the extinction curve. Therefore our spiral knots being closer to the starburst relation and likely having an extinction law comparable to Calzetti et al. (see previous section) is consistent with this conclusion. However, we should keep in mind that in the previous section we assumed a constant $\beta_o$ (i.e. dust-free UV slope). Variations in $\beta_o$, such as with higher contributions of older stars would affect this as well. However, with the inferred SFRs and intensities of this region, there should be plenty of young stars to dominate the UV emission.

\section{Discussion}

One of the principle questions we address in this paper is the global SFR and obscuration of NGC2442. The only previous measurement of the SFR in NGC 2442 was presented in \citet{rd94}, who find 4.9\msun/yr globally. This already has dust correction of $A_{0.65}$\,=\,0.45 applied to it.  The foreground extinction value is $A_{0.65}$\,=\,0.54 (NED). Using the $H_{\alpha}$ luminosity as given in \citet{mihos}), and updating the luminosity distance and SFR relation \citep{kennicutt98}, as well as the difference in foreground extinction values, we find SFR(H$\alpha$)=4.6\,$\rm{M_{\sun}yr^{-1}}$.  Using UV, 8\um, 24\um, and composite (UV+24) estimates, we find values of 6\,--\,11\msun/yr the latter being the UV-based estimate when corrected for dust using the Meurer relation. We conclude that the true SFR is most likely to lie in the range $\sim$\,6\,--\,7\msun/yr based on a wide range of relations including a HII region-calibrated SFR(24) and  integrated total galaxy light and TIR-calibrated SFR(24), as well as composite (H$\alpha$+24, H$\alpha$+TIR, and FUV+24) relations. This also agrees with the SFR(FUV) if we adopt  the \citet{salim07} relations for both SFR(FUV) and $A_{FUV}$.  All of these relations are calibrated using nearby normal star-forming galaxies, and are therefore applicable to NGC2442. 

 This global agreement between the IR-based and UV-based SFR estimates for NGC2442, requires a steeper UV extinction law -- one in between the SMC and starburst (Calzetti) extinction curves, or between a MW-type and a SMC-type extinction.  However, we do find that the areas of highest SFR intensity are consistent with the starburst extinction law, which may provide a hint for the range of applicability of different extinction laws.  Recently, \citet{siana09} suggested that a steeper UV extinction might be required for at least some high redshift Lyman-break galaxies as well, which highlights the need to understand this uncertainty better locally.  

We find that the SFR  in NGC2442 most likely lies in the range $\sim$\,6\,--\,7\msun/yr, which is $\sim$\,15\,--\,35$\%$ higher than the value of SFR(H$\alpha$) alone. In order for the H$\alpha$ emission to give the same value of SFR as the 24\umm it would have to be obscured by $\rm{A_{H\alpha}}\sim0.3\,--\,0.5$\,magnitudes globally.  The inferred UV obscuration value is $A_{1600}$\,=\,2.0 (based on the Meurer relation) or $A_{1600}$\,=\,1.5 (required to make the UV and 24\um\ SFR estimates agree).  Based on the \citet{draine03} $R_V$\,=\,3.1 MW extinction curve, we find $A_{6563}$/$A_{1600}$\,=\,0.30,  implying $A_{6563}$\,=\,0.4\,--\,0.6. If we adopt the theoretical SMC extinction curve from \citet{wd01}, we find  $A_{6563}$/$A_{1600}$\,=\,0.14 implying $A_{6563}$\,$\sim$\,0.2\,--\,0.3. An H$\alpha$ extinction of  $\rm{A_{H\alpha}}\sim0.3\,-\,0.5$ required to reconcile our H$\alpha$ SFR with our IR-based estimates is apparently in between these two values. However, if we consider differential attenuation between the stellar continuum and the ionized gas, we have $A_{H\alpha}$\,$\sim$\,2.2A$_{6563}$ \citep{calzetti01}, implying that UV/optical extinction curve in NGC2442 may be closer to the SMC after all. Note however that the uncertainties on all of these estimates are substantial and even more fundamentally that we have assumed a very simplistic geometry, in that everything is treated as a homogeneous dust screen. 

\begin{figure}[h]
\begin{center}
\plotone{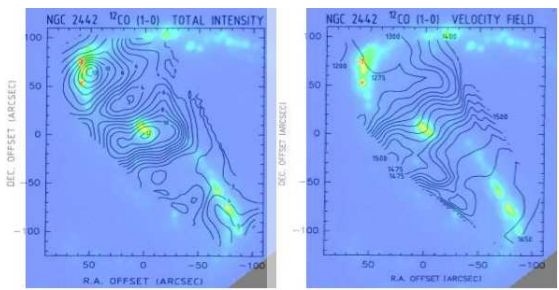}
\end{center}
\caption{Here we overlay our MIPS24\um\ image (i.e. SFR) with the CO total intensity and velocity field images from \citet{bajaja95}. The numbers in the right-hand panel correspond to the velocity contours in km/s. Note that the alignment between the background image and the contour images is done by eye and hence is only approximately correct.  It is clear however that the strong SFR we observe in the northern arm is fueled by significant molecular gas. See \citet{bajaja95} for details on the CO observations. In this figure North is up and East is left. \label{fig_co}}
\end{figure}

Another aspect of this work was to address the spatial distribution of star-formation and obscuration in NGC2442 especially in light of its tidally distorted morphology.  The IR emission (both 8\um\ and 24\um) is fairly clumpy and largely concentrated along the spiral arms, with the notable exception of the superbubble region. The UV emission does not correlate well with the IR emission in that there is relatively stronger FUV emission outside the spiral arms. The superbubble region and the tidal region at the tip of the southern arm are both particularly strong in the UV relative to the IR. We find that the highest SFR intensities are found in the 24\um-bright knots along the spiral arms, followed by the nuclear region which includes a circumnuclear ring. The nuclear region also has the highest levels of UV obscuration with $A_{1600}$\,$\sim$\,4.  The most unusual SEDs are observed for the superbubble region as well as the tidal region, which also has IRAC colors that deviate from the usual trend. The superbubble region in particular, is believed to be an area of strong past and current star-formation outside of the spiral arms. We can also compare the H$\alpha$ image from \citet{mihos} with that of the other star formation tracers.  \citet{mihos} find that there is little diffuse H$\alpha$ emission, with a majority concentrated in H{\sc II} regions.  The most luminous 24\umm and 8\umm regions tend to overlap with the H{\sc II} regions (this can also be seen in Figure\,\ref{fig_ha_8um} for part of the galaxy).  Figure\,\ref{fig_co} also shows a {\it by eye} comparison of the $^{12}$CO (1-0) line intensity and velocity field as determined by \citet{bajaja95} with our MIPS24\um\ map of NGC2442.  Most clearly, the CO intensity exhibits two prominent peaks which correspond to the nucleus and the knots along the northern spiral arm. Both region were identified above as having the highest SFR intensity.  The star-forming knots along the southern spiral arm are somewhat weaker but do correspond to CO-bright regions as well.  Our superbubble region also  shows CO emission as expected for an active star-forming region.    

Lastly, what does this mean in terms of the history of NGC2442?  It is still unclear whether NGC2442 has undergone a tidal encounter with a neighboring galaxy or has suffered ram pressure stripping.  A large cloud of hydrogen, a third the mass of NGC2442's neutral gas, resides just north of the galaxy, and may be a result of these violent processes \citep{ryder01}.  Our best estimate for the total SFR in NGC2442 (SFR$\sim$\,6\,--\,7\msun/yr), is somewhat higher than typical in normal star-forming galaxies \citep[see e.g.][]{smith}, suggesting enhanced star-formation in addition to the morphological distortion \citep[contrary to][]{mihos}. The star-formation is not overwhelmingly concentrated in the nuclear region, as expected in interacting galaxies \citep{smith}, but is rather strong in clumps of intense star-formation along the spiral arms. These regions in themselves display many properties of starburst galaxies including a strong PAH/stars ratio, and potentially a more starburst-like extinction law. The northern spiral arm is  stronger and much more pronounced than its southern counterpart, and includes the clumps of highest star-formation rate intensity, coincident with high molecular gas concentrations (Figure\,\ref{fig_co}). In contrast, the southern part of the galaxy shows a weaker spiral arm which transitions into our `tidal region', as well as the superbubble region outside the spiral arm, which we associate with active star-formation. This north-south asymmetry is rather convincingly explained by ram pressure stripping by \citet{ryder01} where the morphology and star-formation strength of the northern arm is due to the bow shock where the galaxy plows into the intergalactic medium (see also Figure\,\ref{fig_co}). The area of our superbubble region is a strong radio emitter \citep{harnett}, in support of the role of supernovae in that region. While the above discussion generally supports the ram pressure stripping scenario, the details are not yet clear and further study is needed to truly explore the possibilities.  In particular, we need to explore how common this type of galaxy might be at higher redshifts where we generally lack morphological information and where we expect typically denser environments.  Detailed dynamical modeling of NGC2442 would be especially useful for exploring NGC2442's past and explaining the many interesting properties of this galaxy. The morphological and SED properties of the superbubble and tidal regions are unusual and unexpected on the basis of existing simulatations of NGC2442 \citep{mihos}. Higher resolution radio observations of the superbubble region would help shed light on its nature (in particular to see if the shell-like structure observed in the IRAC8\um\ image is visible there as well). A mid-IR spectrum of the tidal region would help us to better understand its peculiar IRAC colors.

Some caveats in our discussion need to be taken into account. While we were careful to use SFR relations that are applicable to our galaxy and to characterize the associated uncertainties, ultimately our conclusions regarding the SFR globally and per region are {\it relative} to the original samples on which these relations were calibrated as well as any underlying assumptions (such as inherent in the H$\alpha$-SFR conversion). In addition, this galaxy has substantial foreground extinction (see Section\,2.4). We correct for this prior to any subsequent analysis, but how would our conclusions be affected by uncertainties in the level of this foreground? Ultimately, our best estimates of the star-formation in this galaxy arise from the infrared data and hence would be unchanged. The SFR(FUV) estimates are strongly affected by dust obscuration and hence, in practice, we use them primarily as a way to constrain the extinction curve in NGC2442. Most simply put, a higher foreground would imply a more SMC-like extinction curve, while a lower foreground would imply a more MW-type dust.  It is difficult to address this further however, because of the large systematic uncertainties on SFR(FUV).  

\section{Conclusions}

NGC 2442 is a nearby galaxy whose peculiar morphology is believed to arise either from tidal interaction with a neighboring galaxy or ram pressure stripping which acts roughly south-to-north.  The {\sl GALEX} and {\sl Spitzer} data allow us to explore the distribution and levels of star formation and dust obscuration in this galaxy.  Our major regions of interest are the spiral arms, spiral knots of high 24\umm emission, nuclear region (including a circumnuclear ring), the diffuse region, a tidally distorted region at the southern tip of the galaxy and a superbubble region inside the diffuse area.  We employ various tracers of star formation: the FUV, 8\um\ and 24\um\ emission as well as a composite FUV+24 SFR relation. For the total SFR, we also consider the $H_{\rm{\alpha}}$ and total IR luminosities. The  dust obscuration is traced using  the UV-slope as well as by trying to reconcile the UV, H$_{\alpha}$ and 24\um-based SFR estimates.  We find that: 
\begin{enumerate}
\item Our best estimate for the total SFR of NGC2442 is $\sim$\,6--7\msun/yr. This is based on the 24\um\ emission as well as a combination of FUV and 24\umm and H${\alpha}$ and total IR emission, as calibrated by \citet{kennicutt09}. 
\item The highest intensity star-forming regions are the spiral knots followed by the nuclear region.  The nuclear region is the most obscured region of the galaxy. It also includes a $\sim$\,0.8\,kpc radius circum-nuclear ring that is especially prominent at 8\um\ (i.e. PAH emission). The nuclear  H${\alpha}$ emission shows a vague extended spiral structure that is largely contained inside this PAH ring. 
\item A tidal debris region has unusual IRAC colors, which is consistent with similar findings from \citet{smith}. With only broad-band data, we cannot say conclusively what is the cause for this, but it may be due to relatively higher levels of hot dust, or the presence of strong molecular or ionic emission lines, especially in the 5.8\um\ band. 
\item With standard UV-slope dust correction,  the SFR(FUV) values are consistently higher than the IR-based estimates.  We find that, globally, we can reconcile the FUV and 24\um\ SFR estimates by a UV extinction law that is intermediate in slope between an SMC-like extinction and a Calzetti-like extinction.  Regions of high star-formation intensity, the spiral knots and the North spiral arm, are consistent with the starburst-calibrated Calzetti extinction.  A similar analysis to reconcile the H$\alpha$ and UV-based SFRs suggests an optical/UV extinction slope intermediate between a MW-type and an SMC-type extinction.
\item We find what looks like a superbubble, $\sim$\,1.7\,kpc across outside the spiral arms of NGC2442. This structure is particularly prominent in the IRAC 8\um\ image, although it can be seen in all of our IRAC images. After the spiral knots, the superbubble has the highest PAH-to-stars ratio, implying significant star-formation. H$_{\alpha}$ and UV emission suggest significant star-formation in the area as well. SN1999ga is found at the edge of this structure.
\end{enumerate}

\acknowledgements
We are grateful to the anonymous referee for their careful reading of our manuscript and  thoughtful suggestions that have significantly improved the content and clarity of this paper.
We would like to thank Beth Willman and Brian Siana for helpful discussions.  Support for this work was provided by NASA through an award issued by JPL/Caltech. This work is based in part on observations made with the {\sl Spitzer} Space Telescope, which is operated by the Jet Propulsion Laboratory, California Institute of Technology under a contract with NASA.  This work uses observations made by Galaxy Evolution Explorer (GALEX), a NASA Small Explorer, launched in 2003 April, and developed in cooperation with the Centre National d'Etudes Spatiales of France and the Korean Ministry of Science and Technology.  This work makes use of data products from the Two Micron All Sky Survey, which is a joint project of the University of Massachusetts and the Infrared Processing and Analysis Center/California Institute of Technology, funded by the National Aeronautics and Space Administration and the National Science Foundation.  This work has also made use of the NASA/IPAC Extragalactic Database (NED) which is operated by the Jet Propulsion Laboratory, California Institute of Technology, under contract with the National Aeronautics and Space Administration.

\clearpage

\begin{deluxetable}{lcccc}
\tabletypesize{\scriptsize}
\tablecaption{Mask Region and Total Galaxy Luminosity Data \label{table_masks1}}
\tablewidth{0pt}
\tablehead{ 
\colhead{Region\tablenotemark{1}} &
\colhead{Pixels} &
\colhead{log($\rm{L(FUV)}$)} &
\colhead{log($\rm{L(8)}$)} &
\colhead{log($\rm{L(24)}$)} \\
\colhead{} & \colhead{} &  \colhead{[$\rm{L_{\sun}}$]} & \colhead{[$\rm{L_{\sun}}$]} & \colhead{[$\rm{L_{\sun}}$]}
}
\startdata
Total Galaxy &  2952 &  9.92 & 10.08 &  9.51 \\
Diffuse Region &   461 &  9.35  & 9.26  & 8.59   \\
North Spiral Arm &   354 &  9.28 &  9.54 &  8.98  \\
Tidal Feature &   189. &  8.68 &  8.43 &  7.89  \\
South Spiral Arm & 166 & 9.11 & 9.29 & 8.70 \\
Spiral Knots &    37 &  8.69 &  9.32 &  8.84  \\
Superbubble &    25. &  8.61 &  8.39 &  7.69  \\
Nuclear Region &    22. &  7.88 &  9.00 &  8.38 
\enddata

\tablenotetext{1}{The aperture used for the total galaxy covers $\rm{1063\,kpc^2}$, corresponding to 2952 pixels of dimension 6\asec$\times$\,6\asec=\,0.6\,kpc\,$\times$\,0.6\,kpc.  }
\end{deluxetable} 

\begin{deluxetable}{lcccc}
\tabletypesize{\scriptsize}
\tablecaption{Obscuration Values \label{table_masks2}}
\tablewidth{0pt}
\tablehead{ 
\colhead{Region} &
\colhead{$\beta\tablenotemark{1}$} &
\colhead{$\rm{A_{1500,Meurer}}$} & \colhead{$\rm{A_{1500,Salim}}$} &
\colhead{$\rm{A_{1500},IR-matched\tablenotemark{2}}$} \\
\colhead{} & \colhead{} &  \colhead{[mag]} & \colhead{[mag]} & \colhead{[mag]}  
}
\startdata
Total Galaxy &   -1.2 & 2.0 &  1.42 & 1.5 \\
Diffuse Region &   -1.3 & 1.9 &  1.34 &  0.6 \\
North Spiral Arm &   -1.2 & 2.0 & 1.42 & 1.8  \\
Tidal Feature &   -1.5 & 1.4 & 0.96 & 0.6  \\
South Spiral Arm &   -1.0 & 2.4 & 1.73 & 1.4  \\
Spiral Knots &   -1.0 & 2.4 &  1.73 & 2.9  \\
Superbubble &   -1.5 & 1.4 & 0.96 & 0.2   \\
Nuclear Region &   -0.01 & 4.4 & 3.28 & 3.5    
\enddata

\tablenotetext{1}{The UV-slope (see \S\,\ref{sect_obsc}).}  
\tablenotetext{2}{The UV obscuration required to match the UV and IR SFRs.  Here we adopt the SFR(24)$_{\rm{HII}}$ for the IR SFR. 
}
\end{deluxetable}

\begin{deluxetable}{lccccccc}
\tabletypesize{\scriptsize}
\tablecaption{Star Formation Rate Estimates \label{table_masks3}}
\tablewidth{0pt}
\tablehead{ 
\colhead{Region} & 
\colhead{SFR(FUV)} &
\colhead{SFR(FUV)-corr\tablenotemark{1}} &
\colhead{SFR(8)} &
\colhead{${\rm SFR(24)_{global}}$\tablenotemark{2}} &
\colhead{${\rm SFR(24)_{HII}}$} &
\colhead{${\rm SFR(FUV+24)}$}  &
\colhead{${\rm SFR(FUV+24)}$ Intensity}  \\
\colhead{} & \colhead{[$\rm{M_{\sun}yr^{-1}}$]} & \colhead{[$\rm{M_{\sun}yr^{-1}}$]} & \colhead{[$\rm{M_{\sun}yr^{-1}}$]} & \colhead{[$\rm{M_{\sun}yr^{-1}}$]} & \colhead{[$\rm{M_{\sun}yr^{-1}}$]} & \colhead{[$\rm{M_{\sun}yr^{-1}}$]} & \colhead{[$\rm{M_{\sun}yr^{-1}kpc^{-2}}$]} 
}
\startdata
Total Galaxy     &   1.7$\pm$1.2   &  10.9$\pm$2.8$\pm$3.8    &   7.6$\pm$1.9    &  4.7$\pm$3.4      &   7.4$\pm$3.4    &  5.9         & 0.006  \\
Diffuse Region   &   0.5$\pm$0.3   &   2.7$\pm$0.7$\pm$0.9    &   1.2$\pm$0.1     &  0.7$\pm$0.5      &   1.1$\pm$0.5    &   1.0       & 0.006  \\
North Spiral Arm &   0.4$\pm$0.3   &   2.5$\pm$0.7$\pm$0.9    &   2.2$\pm$0.5     &  1.4$\pm$1.0      &   2.1$\pm$1.0    &    1.6      & 0.013 \\
Tidal Feature    &   0.10$\pm$0.07  &   0.4$\pm$0.1$\pm$0.1   &   0.17$\pm$0.02   &  0.15$\pm$0.12      &  0.3$\pm$0.12   &  0.2      & 0.003  \\
South Spiral Arm &   0.3$\pm$0.2   &   2.4$\pm$0.6$\pm$0.8    &   1.2$\pm$0.3     &   0.8$\pm$0.6     &   1.1$\pm$0.5    &  0.9        & 0.016  \\
Spiral Knots     &   0.10$\pm$0.07  &   0.9$\pm$0.2$\pm$0.3   &   1.3$\pm$0.8      &  0.8$\pm$0.6      &   1.2$\pm$0.6    &  1.0       & 0.076  \\
Superbubble      &   0.09$\pm$0.06  &   0.3$\pm$0.1$\pm$0.1   &   0.15$\pm$0.03    &  0.08$\pm$0.06      &  0.12$\pm$0.06   &  0.2     & 0.017 \\
Nuclear Region   &  0.02$\pm$0.01 &   0.9$\pm$0.2$\pm$0.3     &   0.6$\pm$0.4     &  0.3$\pm$0.2      &   0.4$\pm$0.2    & 0.3         & 0.041
\enddata

\tablenotetext{1}{SFR(FUV)-cor is corrected for dust obscuration by multiplying the uncorrected star formation rate by $\rm{e^{A_{1600}/1.086}}$.  The uncertainty in SFR(FUV) is from the spread in SFR(FUV) relations in the literature, while the uncertainty in SFR(FUV)-corr is from the uncertainty in the UV-slope dust correction (the first number), and the spread in the SFR(FUV) relations (the second number). }
\tablenotetext{2}{${\rm SFR(24)_{global}}$ is calculate using a non-linear relation calibrated using integrated normal star forming galaxies by \citet{zhu08} and given in \citet{calzetti10}, while ${\rm SFR(24)_{HII}}$ is calculated using a non-linear relation calibrated using HII regions by \citet{relano07}.}

\end{deluxetable} 

\clearpage

\appendix
\section{Further details on the nuclear region}

As mentioned in the Introduction, the nucleus of NGC2442 is a LINER. The optical spectra (Bajaja et al. 1999) covering from $\sim$\,3500\,--\,7000\AA, shows some of the
classical emission lines including: H$\alpha$, H$\beta$, [OII] 3727, [OIII] 4959+5007, [NII]5755+6548+6584, [OI] 6300 (surprisngly not 6363) and [SII] 6717+6731. The fact that there is a significant energy source exciting the gas can be seen in that, for example, the optical [OIII] 5007 is as bright as H$\beta$. 

\begin{figure}[h]
\begin{center}
\epsfig{file=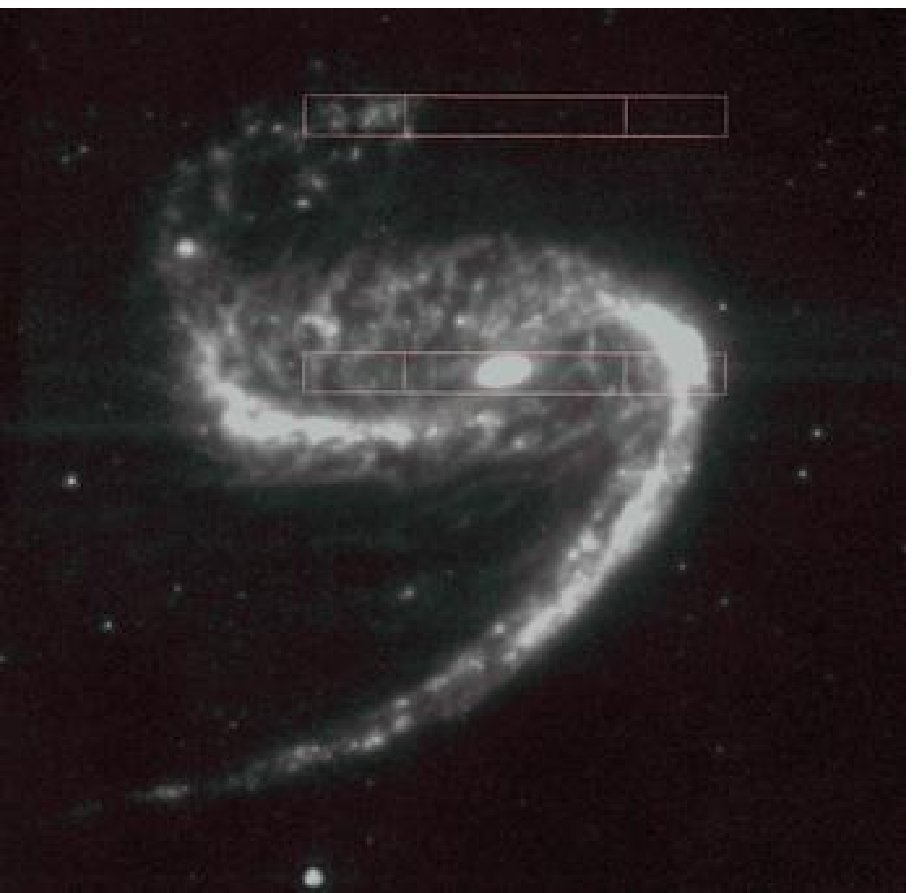, clip=, width=7cm}
\epsfig{file=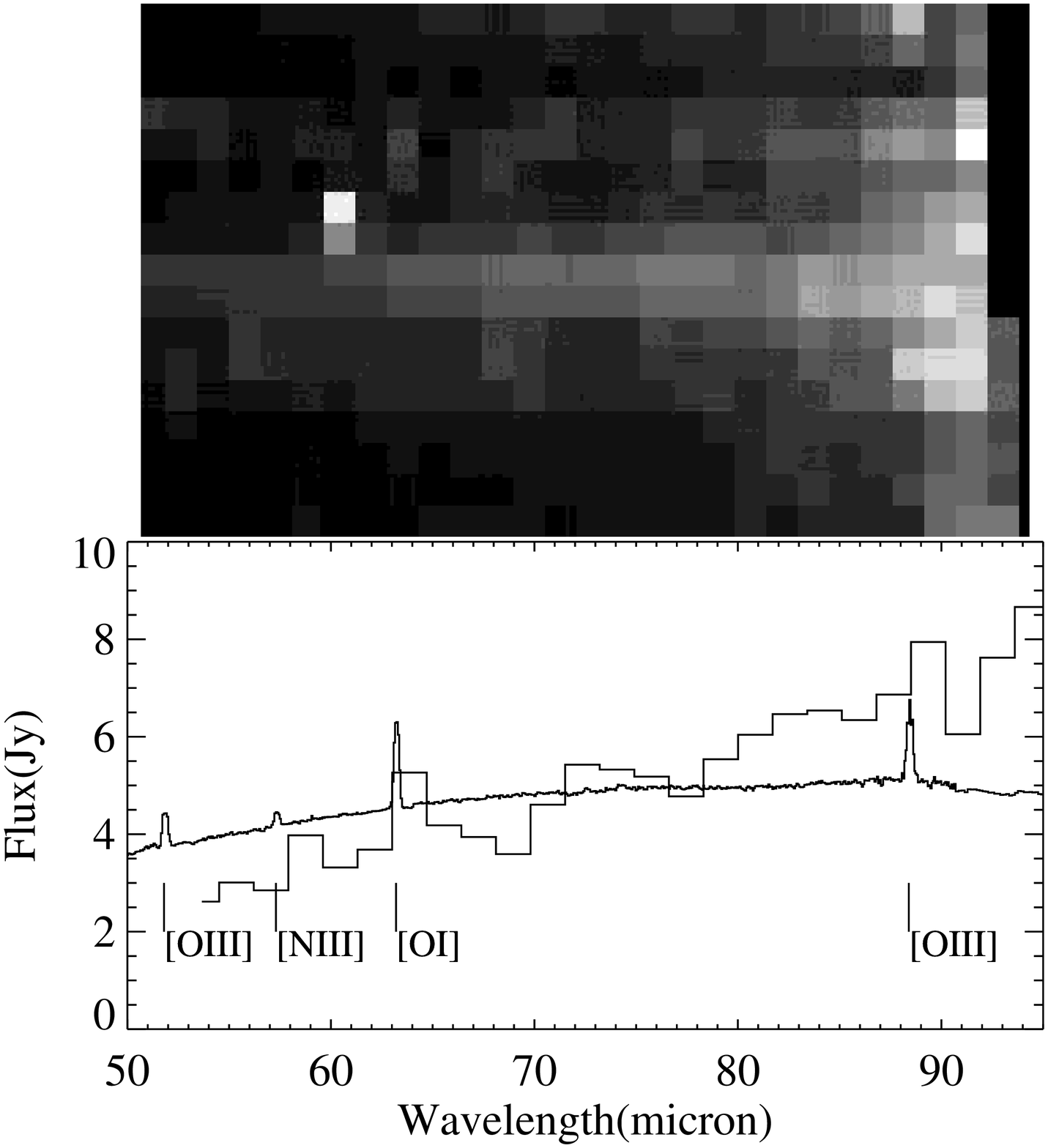, clip=, width=7.0cm}
\end{center}
\caption{{\it Left:} The footprint of the MIPS SED observation overlaid on the IRAC 8\um\ image. {\it Right:} The 2D and 1D low resolution MIPS SED-mode spectrum of the nuclear region of NGC2442. For  comparison we also show the {\sl ISO} spectrum of M82 \citep{colbert99}. Two features to note are that the NGC2442 nuclear spectrum is fairly steeply rising, and that it seems to contain a strong oxygen 63\um\ line. Note however that the oxygen line is offset from the continuum by roughly $\sim$\,6\,--\,15\asec. This suggests that while the observed continuum is likely associated with the nucleus itself, the oxygen line is more likely associated with the star-forming nuclear ring. \label{fig_nucleus_sed}}
\end{figure}

Although, we found that the nuclear IRAC colors were not in themselves indicative of an AGN, this is actually not surprising for LINERS. For example, FRI radio galaxies often show no IR
emission from their AGN \citep{ogle06}. There is an on-off MIPS SED observation of the nuclear region of NGC2442. This observation was made Feb 20, 2004 (on the same day as the IRAC Chan4 image). The MIPS-SED FOV is $\sim$\,2.7\,$\times$\,0.34\amin\, which covers the whole nuclear region (including the 8\um\ ring), but it is also slightly contaminated by the edge of the spiral arm (see Figure\,\ref{fig_nucleus_sed}{\it left}). In Figure\,\ref{fig_nucleus_sed}{\it right}, we show both the 2D and 1D MIPS SED, which is compared with the well known spectrum of M82. The [OI] 63\um\ line that we can see in the low resolution SED spectrum appears to be offset from the nucleus itself as can be seen in the 2D spectrum. It appears that the continuum likely originates in the nucleus itself whereas the oxygen emission is more likely associated with the ring. It is interesting that the continuum is rising compared with M82, which is rather `warm' for a starburst. As was already discussed in the context of the nucleus IRAC colors, this  also suggests the lack of a hot AGN-heated torus. 
 
\begin{figure}[h]
\begin{center}
\epsfig{file=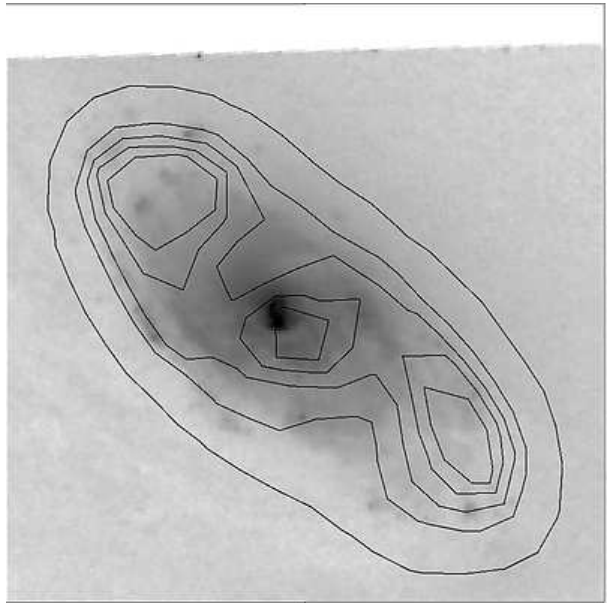, clip=, width=8cm}
\end{center}
\caption{Here the greyscale shows the archival ACS H$\alpha$ image with the IRAC 8\um\ contours overlaid. The box is 20\asec\,$\times$\,20\asec.  North is up and East is left. \label{fig_nucleus_ha}}
\end{figure}

The other interesting feature in the nuclear region is that in the IRAC images (especially at 8\um) it can be separated into a compact nucleus and a ring of a radius of $\sim$\,0.8\,kpc. In Figure\,\ref{fig_nucleus_ha}, we compare the 8\um\ emission (contours) with an archival ACS H$\alpha$ image available through the MAST {\sl HST} archive. The conclusions that can be drawn from this comparison are: 1) the H$\alpha$ emission is largely inside the 8\um\ ring meaning the latter can be interpreted as a higher obscuration/dustier region whose PAH are excited by the stars inside the ring; 2) the H$\alpha$ and 8\um\ nuclei are slightly offset from each other, which could be due to obscuration, although some pointing offset cannot be excluded; and 3) the H$\alpha$ nucleus is slightly resolved and there is a suggestion of spiral structure in the H$\alpha$ image which type of structure has been seen in other star-forming galaxies, this supports the view that although an AGN is likely present (accounting for the LINER spectrum), significant nuclear star-formation is also present. The strong (and rising) far-IR continuum seen in Figure\,\ref{fig_nucleus_sed}{\it right} also supports this view.

\end{document}